# Optimizing Climate Policy through C-ROADS and En-ROADS Analysis

By: Iveena Mukherjee[1*]
[1]The Charter School of Wilmington, Wilmington, DE
*Corresponding author: mukherjee.iveena@charterschool.org

## Abstract

With the onset of climate change and the increasing need for effective policies, a multilateral approach is needed to make an impact on the growing threats facing the environment. Through the use of systematic analysis by way of C-ROADS and En-ROADS, numerous scenarios have been simulated to shed light on the most imperative policy factors to mitigate climate change. Within C-ROADS, it was determined that the impacts of the shrinking ice-albedo effect on global temperatures is significant, however differential sea ice melting between the poles may not impact human dwellings, as all regions are impacted by sea ice melt. Flood risks are also becoming more imminent, specifically in high population density areas. In terms of afforestation, China is the emerging leader, and if other countries follow suit, this can incur substantial dividends. Upon conducting a comprehensive analysis of global trends through En-ROADS, intriguing patterns appear between the length of a policy initiative, and its effectiveness. Quick policies with gradual increases in taxation proved successful. Government intervention was also favorable, however an optimized model is presented, with moderate subsidization of renewable energy. Through this systematic analysis of assumptions and policy for effective climate change mitigation efforts, an optimized, economically-favorable solution arises.

## 1. Introduction

As climate change continues to shape many facets of our world, climate simulations become an important tool to model the effects of policies on societies to account for any negative externalities, before they occur. With the United Nations calling for a 43% reduction in global greenhouse gas emissions by 2025 and net zero emissions by 2050 to "limit warming to 1.5° Celsius above pre-industrial levels," it becomes evident that "current national commitments are not sufficient" to meet the outlined goals (United Nations, 2022). Thus, it becomes necessary to outline efficient and feasible steps that can be taken to mitigate the impacts of climate change.

When analyzing the different remediation strategies for climate change and its impacts, the importance of different factors becomes essential to a cost effective approach that minimizes deadweight loss and encourages further growth. Past studies on utilizing climate simulations to foster growth and understanding around policy have proven successful, especially in increasing the urgency of public action (Hensel et al., 2023). Apart from this public setting, climate simulations such as En-ROADS have also proven helpful in corporate settings, to increase employee understanding of their role and tasks with risk factor mitigation regarding climate policies within companies (Kapmeier et al. 2021).

However, an important application of climate modeling to assist governments in strategic outcomes from policy has yet to be explored. As Betts (2005) noted, a thoroughly integrated approach to climate crop modeling is needed, and, in conjunction with government policies, this may yield positive results when put to the test of climate change mitigation.

As it stands, collaboration across borders on policy initiatives is a fascinating avenue for increased climate change reduction dividends, and one which must be analyzed for future policy. En-ROADS and C-ROADS facilitate this relationship between simultaneous climate policies, as each allows multiple countries' policies to be simulated at once. Additionally, through the use of data from the "Land Use Harmonization (LUH2) data prepared for the Climate Research Program Coupled Model Intercomparison Project (CMIP6)," data amongst factors can be synthesized to create the most comprehensive model available (C-ROADS, 2023; En-ROADS, 2023).

Domestic policy in larger countries is also an increasingly important field of study, as regional impacts can lead to diverse effects from the same policy initiative, due to differences in infrastructure, innovation, cost-of-living, industry and a multitude of other factors. Through analyzing several policy factors including the duration of initiatives, the initial assumptions and pace of progress from region to region, policy initiatives can be optimized to best serve citizens with the least economic and social cost amidst different political and social constructs and beliefs.

In the past, studies have been conducted on the "ratcheting effect," in which measures to reduce greenhouse gas emissions were increased periodically. According to Holz et al. (2018), without significant and stringent policies on industry and other aspects of the economy, attaining the previously outlined goal of limiting warming to 1.5° Celsius above pre-industrial warming levels is impossible. This paper seeks to outline a policy scenario that meets the UN climate goal, whilst also maintaining economic prosperity.

Economic prosperity becomes important as environmental justice comes to the forefront. Developing countries and their respective economies are at a greater risk of losing out due to climate change, due to changes in capital and the order of distribution, thus disrupting supply chains and increasing the wealth gap among nations (Fankhauser, 2005). Additionally, when it comes to fluctuations in prices, this can have an impact on human health. Many proposed climate actions increase prices of goods incrementally, and when applied to markets of necessary goods and services, such as the food industry, it has been found that increased food prices can lead to amplified health inequalities, and reduced nutrient density in diets (Lake et al., 2012). Navigating a climate policy which ensures equitable economic impacts for all groups is the most desirable scenario.

Policy factors aside, natural factors can also have a sizable impact on progress towards a greener society. A policy approach which takes into account the changing atmospheric conditions, along with receptibility of the climate to human actions will be most successful. A few key natural factors include climate sensitivity, sea level melt rates, and population growth projections. These natural factors are important to include at different increments, as they can affect the outcome of the policy results. Having this stabilizing feature is helpful, as in past policy negotiations, base amounts of $CO_2$ emissions, population growth projections, and melt rates have been reported differently among different countries, making negotiations difficult to navigate (Sterman et al., 2012). Being able to change and maintain these initial factors increases the likelihood of accurate comparisons and considerations under different climate goals, thus consolidating the body of previous work into more manageable climate action plans.

This paper fills knowledge gaps among the interrelationships between different countries' climate policies regarding peak emission years, afforestation, taxation, subsidies and other policy tools in use. Outlined are the impacts from these comprehensive factors, along with their policy implications, and a specific combination of policies for the most efficient method to meet the United Nations goals, whilst maintaining economic prosperity.

# 2. Methods

## 2.1 C-ROADS

To analyze the impact of individual nations and nation groups on the onset of climate change, C-ROADS was used, with several key metrics in consideration. First, the peak emissions year and first year of carbon dioxide emissions reduction was noted. This was an important factor considered as it is typically a policy that countries declare forthright, and countries have received some guidance from the United Nations on this topic, ie. most countries should have a peak emissions year by 2025 (United Nations, 2022). Next, factors such as annual reduction rate and preventative deforestation rate were accounted for (Sterman et. al, 2012), as these are the goals countries typically set for themselves. In terms of analyzing the results, the measures of efficacy include the temperature increase over the next 8 decades, the $CO_2$ concentration in 8 decades, the sea level rise in 2100, and the cumulative avoided $CO_2$ emissions by 2100, assuming a steady increase without these policy measures. Temperature, $CO_2$ concentration, sea level rise and avoided $CO_2$ emissions were chosen as key indicators as they illustrate the most pertinent impacts on human health and daily life, and have been cited by previous literature as important comparisons in relation to goals set out by the United Nations (Sterman et al., 2012; Holz et al., 2018).

In analyzing the impacts of a country's policy, the nations and groups of nations are grouped into six categories: the United States, the European Union, 'Other Developed Nations,' China, India, and 'Other Developing Nations.' The countries grouped under "developed" and "developing" nations are based on UN nations/blocs of negotiators (Zahar, 2019), with notable countries in each bloc listed in Appendix A. In terms of data collection, C-ROADS obtains its data from a variety of sources, on its different topics. It collects data on $CO_2$ emissions from fossil fuel sources (Friedlingstein et al., 2022), world population prospects (United Nations, 2022), and global GDPs (World Bank, 2022).

C-ROADS is a coagulation of several types of models, along with novel equations to model distinct scenarios. For modeling sea level rise, it uses a semi-empirical model from Vermeer and Rahmstorf (2009) which factors in melt contributions from Greenland and Antarctica, along with contributions from artificial reservoirs. For modeling energy and emission changes, C-ROADS uses a variety of reference calculations, and models the differences in GDP per capita in relation to temperature change, due to the variable nature of this relationship (C-ROADS, 2023). For modeling the carbon cycle impacts, C-ROADS uses a variation of the FREE model (Fiddaman, 1997), with recently calibrated data. This model includes modeling the shift in chemical equilibria of the atmosphere and mixed ocean layer, along with using the eddy-diffusion model for deep-ocean interactions (Oeschger, Siegenthaler et al., 1975). By considering both the basis behind physical changes in the environment in addition to the changes' effects on the economy, C-ROADS provides a comprehensive avenue to model climate change and its vast implications.

## 2.2 En-ROADS

To analyze the impacts of global policy initiatives on the total output of several key metrics, the systematic model En-ROADS was used. En-ROADS has two key features which can be analyzed, namely assumptions in the background and policies in the foreground (Siegel et. al, 2018). To start, modern assumptions were changed slightly, to account for the newer landscape of environmental engineering, and other innovations. Although retrofitting infrastructure can be a positive way to protect against climate change (Liu et al., 2022), the maximum retrofit potential for buildings and industry was changed from 70% to around 50% to account for aging infrastructure that becomes increasingly difficult to retrofit, mitigating opportunities to achieve the same efficiency levels as new construction. Another aspect of assumptions that were modified were the COVID-19 impacts, and the duration of the impacts on consumption. Although the default setting was 1 year, this has since changed to 3 years, as even almost 4 years past the COVID-19 pandemic, markets are still seeing new consumer behaviors emerge in light of the pandemic (Verhoef et al., 2023). In terms of the policies that were modified, a key factor was the length and goal-setting duration of the policy initiative in question, which ranged from 10 to 60 years. Another input was the amount of methane gas produced through agriculture, in light of emerging technologies regarding the feed of cattle and reducing methane production through other impactful methods. Other changes in policy included the taxation and subsidies of energy resources, and the change in reliance on non-renewable energy over time. The price of carbon emissions per ton was also a factor included during this analysis, and one which stands to greatly impact the economy. These policy changes were included in this analysis of plausible climate policies as they represent possible changes governments can make in the future to shape climate policy towards meeting the UN goals.

Similar to C-ROADS, En-ROADS uses the Vermeer and Rahmstorf (2009) model to model sea level rise. For modeling the demand and supply of energy, attractiveness is considered as an exponential function of "cost, complementary assets (for transport only), and other factors including phase-out policies, technical feasibility, and other effects" (En-ROADS, 2023). This measure of attractiveness is then used in conjunction with the energy intensity of new capital, and the long term energy intensity to model supply and demand. For modeling forestry impacts, the type of land is split into four types: forest, agriculture, other, and tundra, with different ages for forests. From here, the carbon flow for each land type is considered, and multiplied by constants to get the resulting GDPs and other impacts. For modeling emissions, emission factor rates are taken from the Food and Agriculture Organization of the United Nations. In terms of data collection and validation, Climate Interactive checks its models against the Integrated Assessment Models used by the Intergovernmental Panel on Climate Change (En-ROADS, 2023).

# 3. Results & Discussion

## 3.1 C-ROADS: Impact of Sea Level Melt

Sea level rise is typically the most common “symptom” of climate change, however when observing the rate at which it can happen, there are several factors which can change the rate of this process. This includes the increased entrapment of greenhouse gasses, and other warming factors, creating a positive feedback loop. Ice has the highest albedo of any widespread cover on the planet, indicating its usefulness in cooling the planet, through absorbing a lower amount of solar energy. With less surface area covered by ice, the overall albedo of the Earth’s surface decreases, thus becoming warmer as a greater amount of solar energy is trapped. Additionally, the darker, lower layers of the ocean tend to absorb more heat further warming the planet, and feeding into the aforementioned positive feedback loop. In C-ROADS, the status quo addition of melt from Antarctica and Greenland is 0.11 m by the year 2100. However, this is a conservative estimate, and when looking at literature (Jevrejeva et al., 2014), the probability of having a melt of greater than 0.18 m is 0.05, indicating that any value slightly below 0.18 would be admissible. This type of melt is most likely to impact metropolitan areas, close to the coast, or other types of bodies of water.

As an example, New York City, and specifically its borough, Manhattan were observed. Although Manhattan and other New York City boroughs remained safer from the levels of increased sea level melt, their counterparts across the Hudson river in New Jersey face increased pressure from the Hackensack River, posing a threat to human life and property. Whilst this is an isolated example, through the C-ROADS prediction model, it is estimated that just over 399 million people will be at a flood risk with this minor increase in sea level melt, a 25 million person increase from the baseline sea level melt increase of 0.11 m. Looking at previous literature regarding the greater New York City region, this result is comparable, as it was predicted that 100 year interval floods in this region would be reduced to 19-68 year interval floods by 2050 (Gornitz et al., 2001).

A more concerning statistic however, is the increase in population density below the high tide sea level. While flooding is a major threat, most floods occur within longer flood intervals, making floods events that occur sparsely throughout the years. With this increase in sea level melt, an additional 24 million people will be living below the high tide line, indicating that dwellings and communities will be affected by the sea level rise incessantly in coming years.

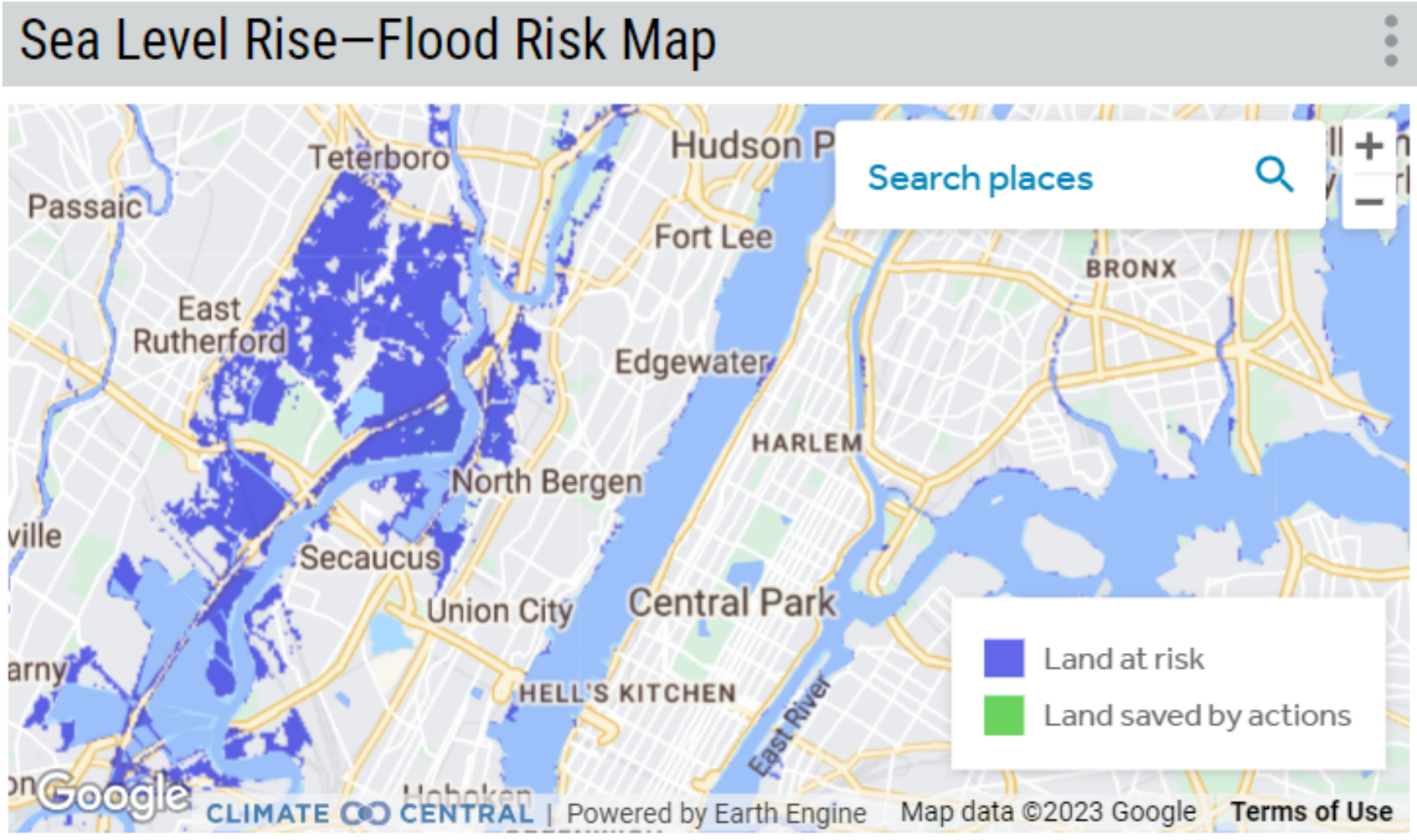

Figure 1: 110 cm. sea level rise around the New York City region

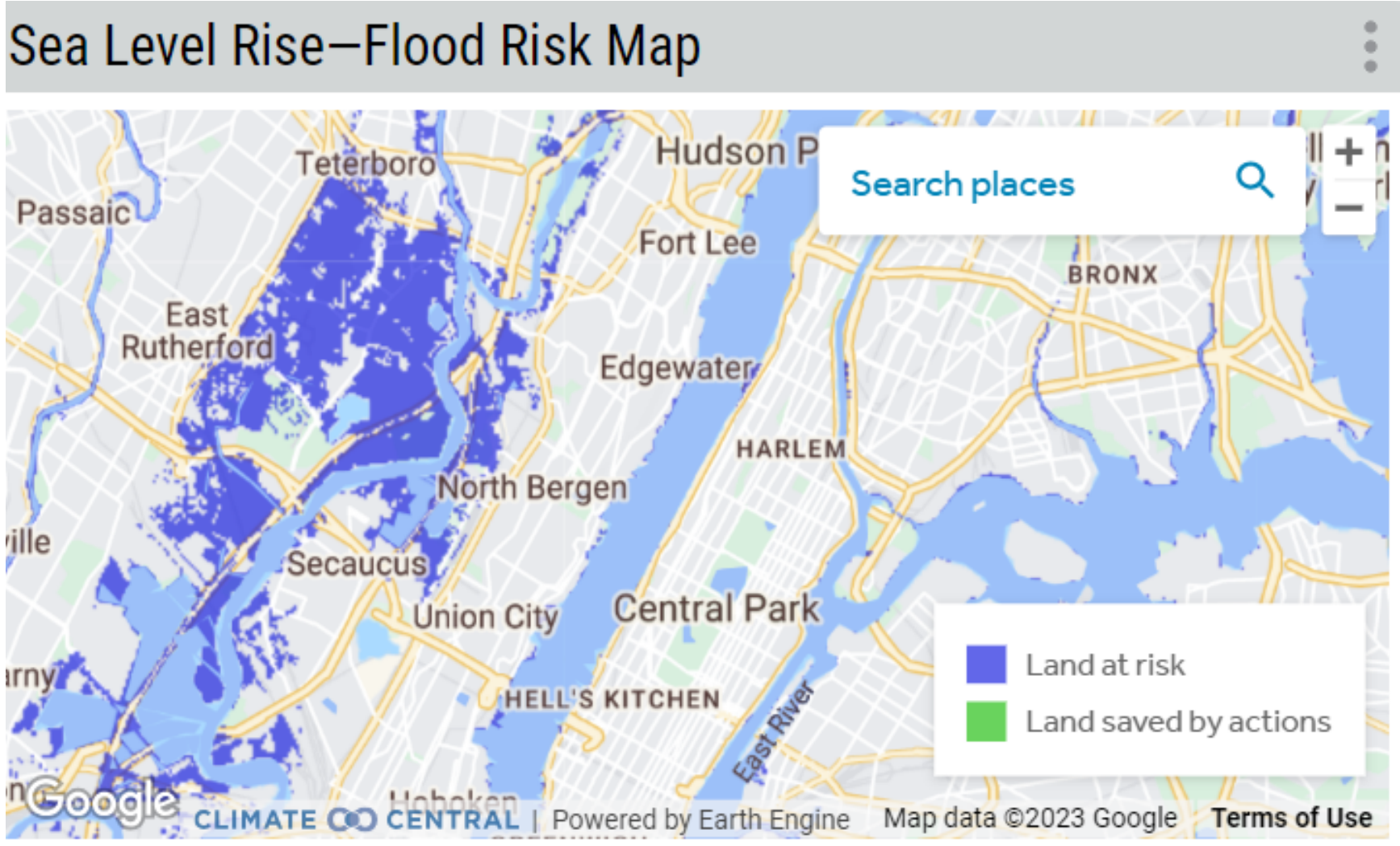

Figure 2: 180 cm. sea level rise around the New York City region

To see if the impacts of the sea level melt were localized, the location of the sea level melt was also considered. For impacts within the Northern Hemisphere, the Greenland Ice Sheet was melted at the higher rate of 0.18 m by 2100, leaving Antarctica at the status quo. Looking at global measures of warming, such as the temperature, it was found that neither ice cap had a greater effect than the other at increasing the positive feedback loop. This emphasizes the need for collective action against climate change, from as many distinct locations as possible.

## 3.2 C-ROADS: Impacts of Economic & Population Growth

Population growth has long been a concern among different countries, but as the impacts of climate change become imminent, it may affect the ability of some countries to transition to climate-friendly processes and infrastructure. To model the impacts of human growth and development on the future of the climate, the status quo of 8.8 billion people was set as the baseline until 2100, assuming equivalent birth and mortality rates. Although this is an unlikely scenario, it was used as a baseline to compare against other scenarios, at the tail end of the UN population range. Coupled with an economic growth increase of 2.5% per year, this brought the global average temperature at the end of the 21st century up by 3.5° Celsius. Looking at the current growth models, with approximately 10.4 billion people by the year 2100, with 2.5% economic growth, this increases the temperature by 3.7° Celsius, only increasing by .2° Celsius more, with an additional 1.6 billion people. Although this is surprising, the statistic with purely 12.4 billion people, and a standard economic growth rate of 1.5% only increases global temperatures in 2100 by 3.4° Celsius. These results emphasize the increasing importance of the economy, and large scale global factors on global warming. The strong and positive correlation between economic growth and increased global temperatures seems to have a larger impact than a more populous society, because of the long-ranging impacts that the economy can have on the environment. As technological advancement becomes more widespread, and life expectancies increase, this can lead to a 4% increase in output (Bloom et al., 2001). As climate economists work to navigate solutions to decrease the use of increasingly carbon-heavy processes, socioeconomic conditions must also be considered, as developing countries are more likely to be at the hands of economic risk, as changing capital and systems can disrupt communities (Fankhauser, 2005) reliant on a single market, or source of income. However, a more equitable distribution of resources, and the increase in economic growth and output can cause environmental solutions to become mainstream and more accessible, thus solving the root cause of the problem. Considering the world population's high estimate by the end of the 21st century is 20 billion individuals (Gilland, 1995), this increase in life expectancy, and thus output and innovation is necessary to undo the impacts of years of industry.

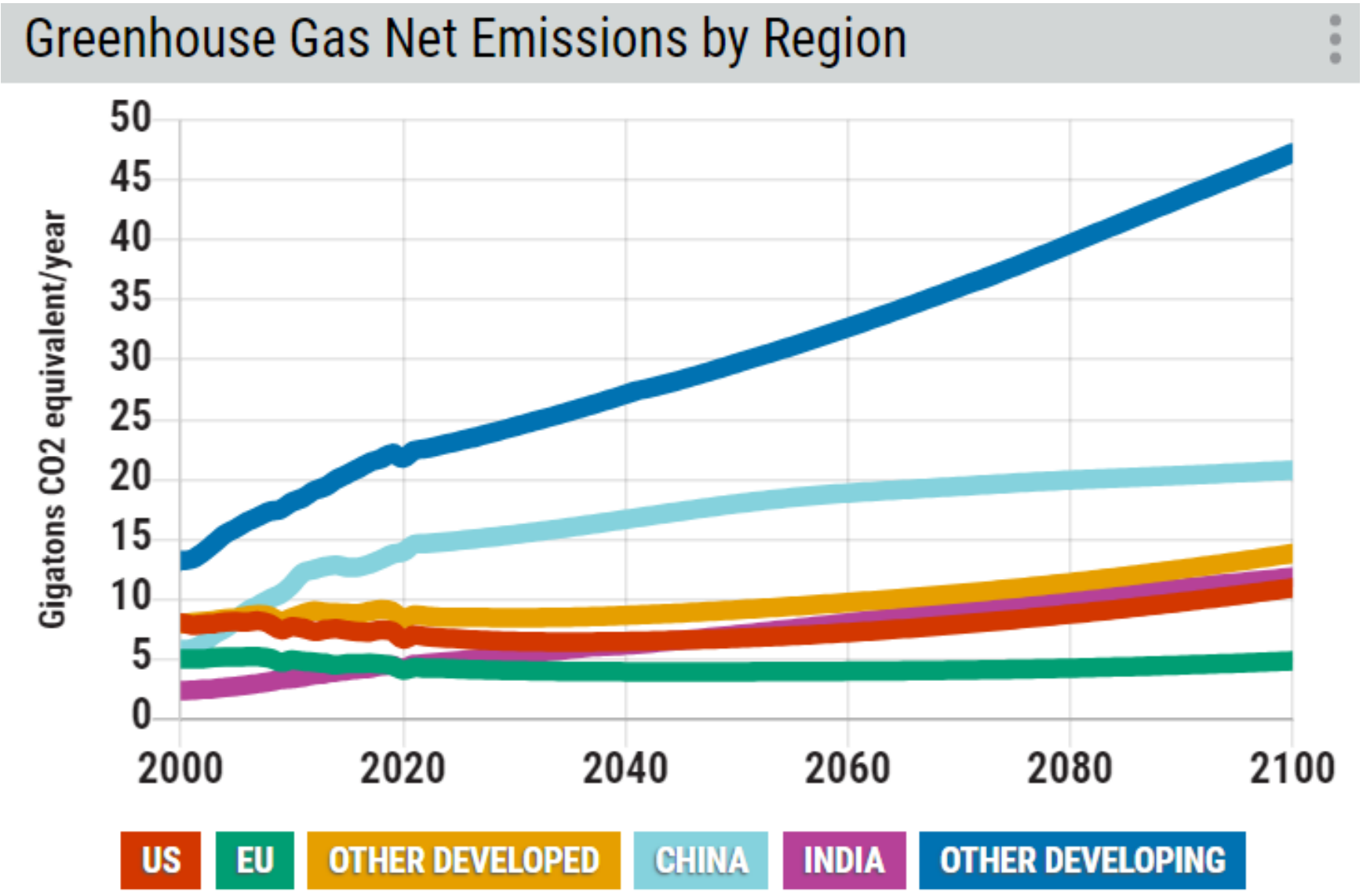

Figure 3: GHG emissions by region with a 2.5% economic growth rate and status quo centennial population projection (10.4 billion)

## 3.3 C-ROADS: Impact of Climate Sensitivity

As climate actions have impacts, the way in which the environment responds to increases in greenhouse gasses (GHGs) and specifically $CO_2$ vastly impacts the future temperatures of the planet. Climate sensitivity is described as the temperature increase per doubling of $CO_2$ in the atmosphere. Throughout literature, there are estimates of climate sensitivity falling around 1.5 - 4.5° Celsius with a high level of confidence (Stocker, 2014). In C-ROADS, the status quo is approximately 3° Celsius. However, increasing the climate sensitivity just slightly to 4.5° Celsius led to an increase in global temperature over the next century of up to 4.6° Celsius, the highest of any factor considered in this study thus far. This indicates the importance of reducing $CO_2$ as one of the main causes behind climate change, as its impacts alone, regardless of population, economic growth, or sea level melt are devastating. In terms of policy impacts, this can mean placing further limits on industrial $CO_2$ production, or promoting the practice of green finance through incentives, which has been shown to have a negative correlation with $CO_2$ emissions (Meo and Karim, 2022).

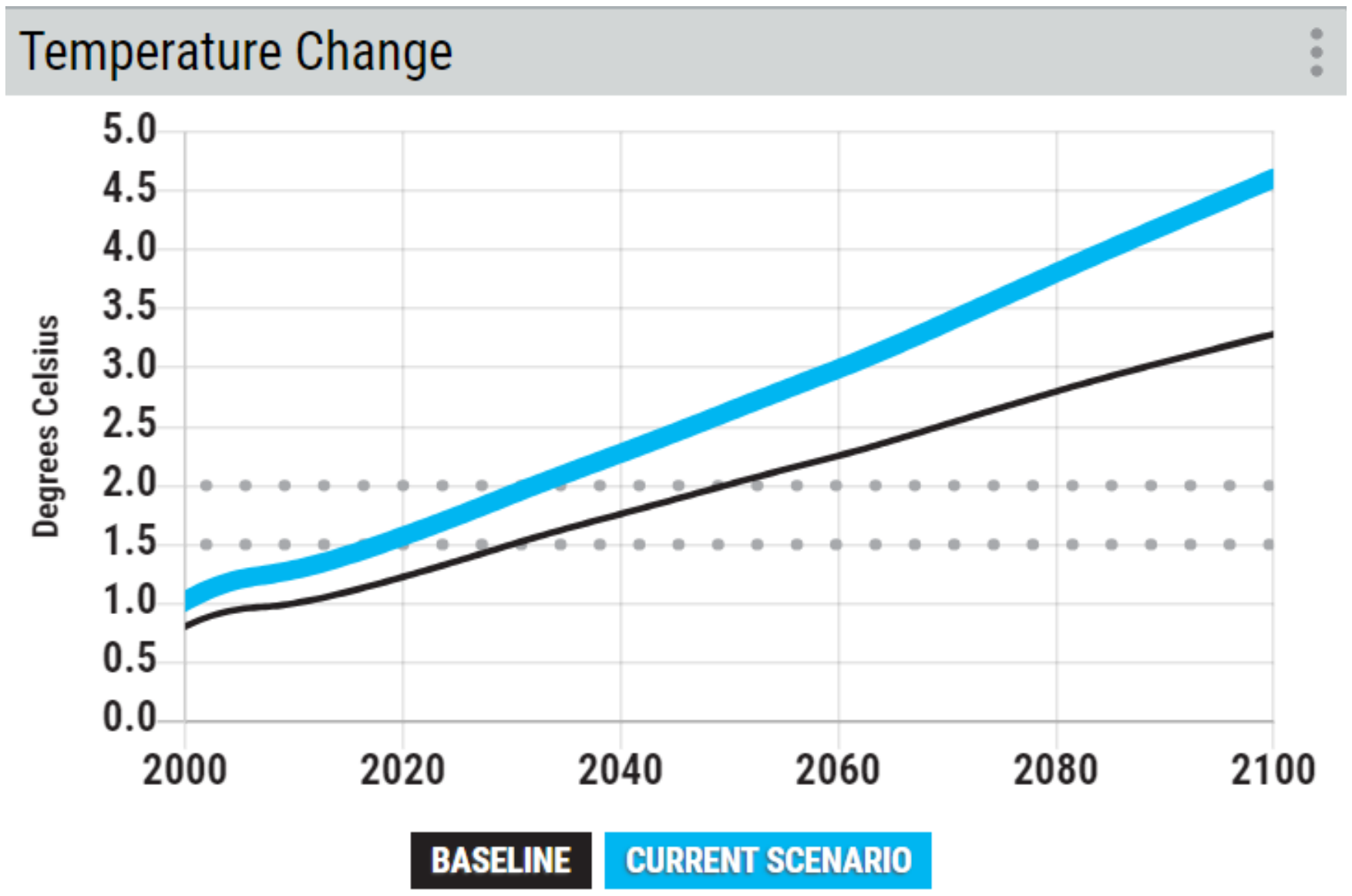

Figure 4: Temperature change with 4.5C climate sensitivity

## 3.4 C-ROADS: Impact of Country-Specific Emission Reduction

When analyzing the global makeup of the climate crisis, there are frequently a few key global players who can work to make progress on the crisis. The United States is frequently brought up as one of the countries who should take strides to cut down on emissions, especially as the economic growth rate and population growth rate are positive. However, upon conducting C-ROADS simulations on the scenario that the United States reduces emissions by a staggering 10% annually starting in 2050 after a spurt of innovation, the global temperature graph over the course of the next century seems abysmal. Even after a 10% decrease in emissions, year after year, for nearly half a century, the predicted global temperature is just 0.1° Celsius cooler, at 3.2° Celsius in 2100.

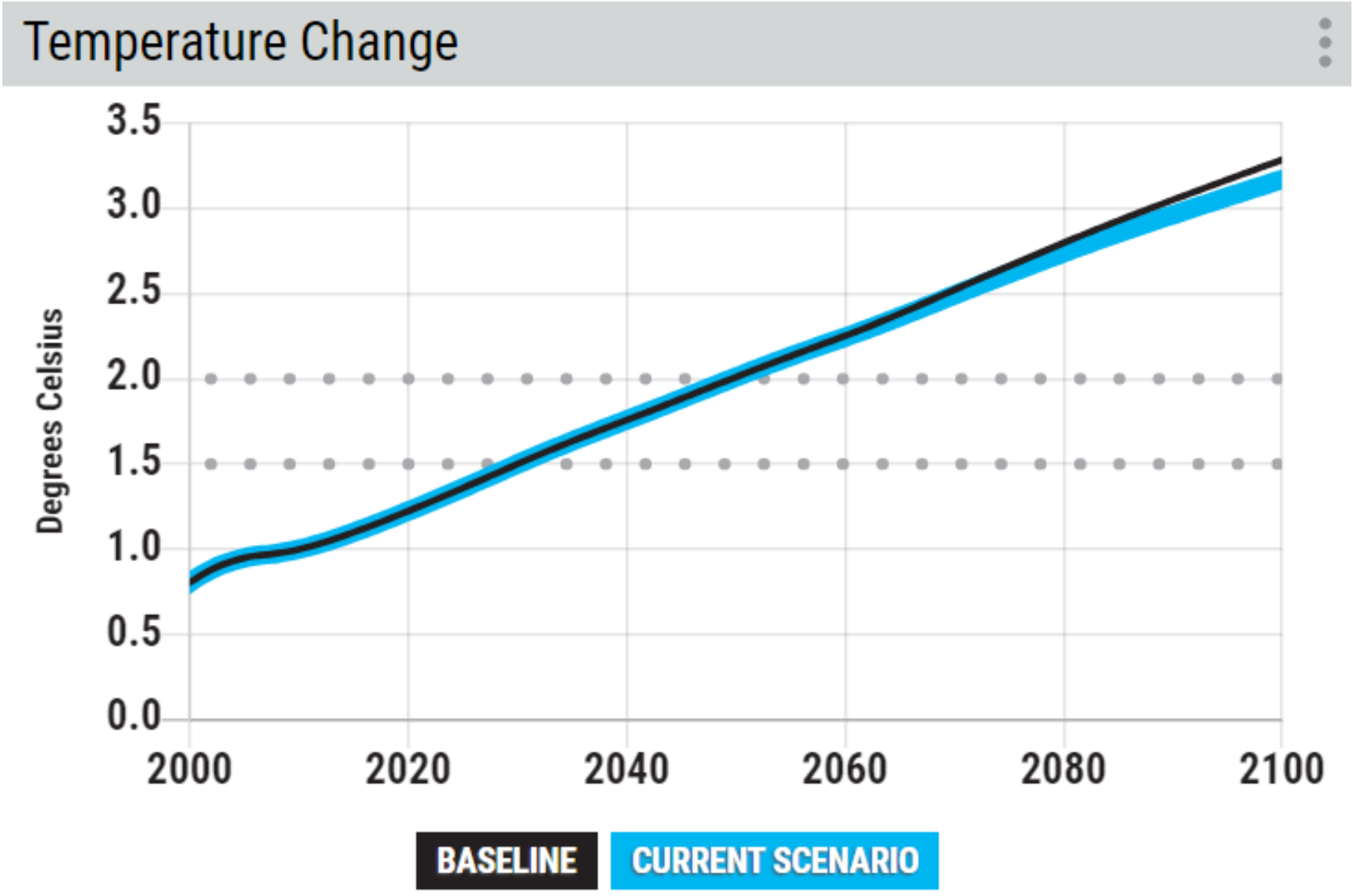

Figure 5: Temperature change with 10% US emissions reduction starting 2050

However, this proves not to be the only solution. If diplomatic ties remain steady, and policy makers encourage the growth of green technology, along with China stepping in to decrease emissions by 50% annually following the year 2050, the future looks promising. By introducing China as another country to reduce global emissions, the global temperature at the end of the century falls a further 0.3° Celsius, coming to 2.9° Celsius. While it may seem beneficial to have just China reduce emissions, coupling the US's processes and decreasing emissions across the Pacific is beneficial, as it promotes a more sustainable trade ecosystem. By introducing both countries as partners against the climate crisis, other countries will soon follow suit, as both financial and manufacturing powerhouses will be starting to embrace sustainability. Recently, the realm of "green finance" has entered markets, and is described as "the financing of public and private green investments" (Lindenberg, 2014). This may prove a powerful tool which governments can leverage through incentives. Although past studies using climate simulations have focused on the United States as the sole conduit for reducing global emissions (McNicholas, 2023), by using the geopolitical influence of either country, both countries can set a sustainable example, which may also encourage other countries to participate in green finance.

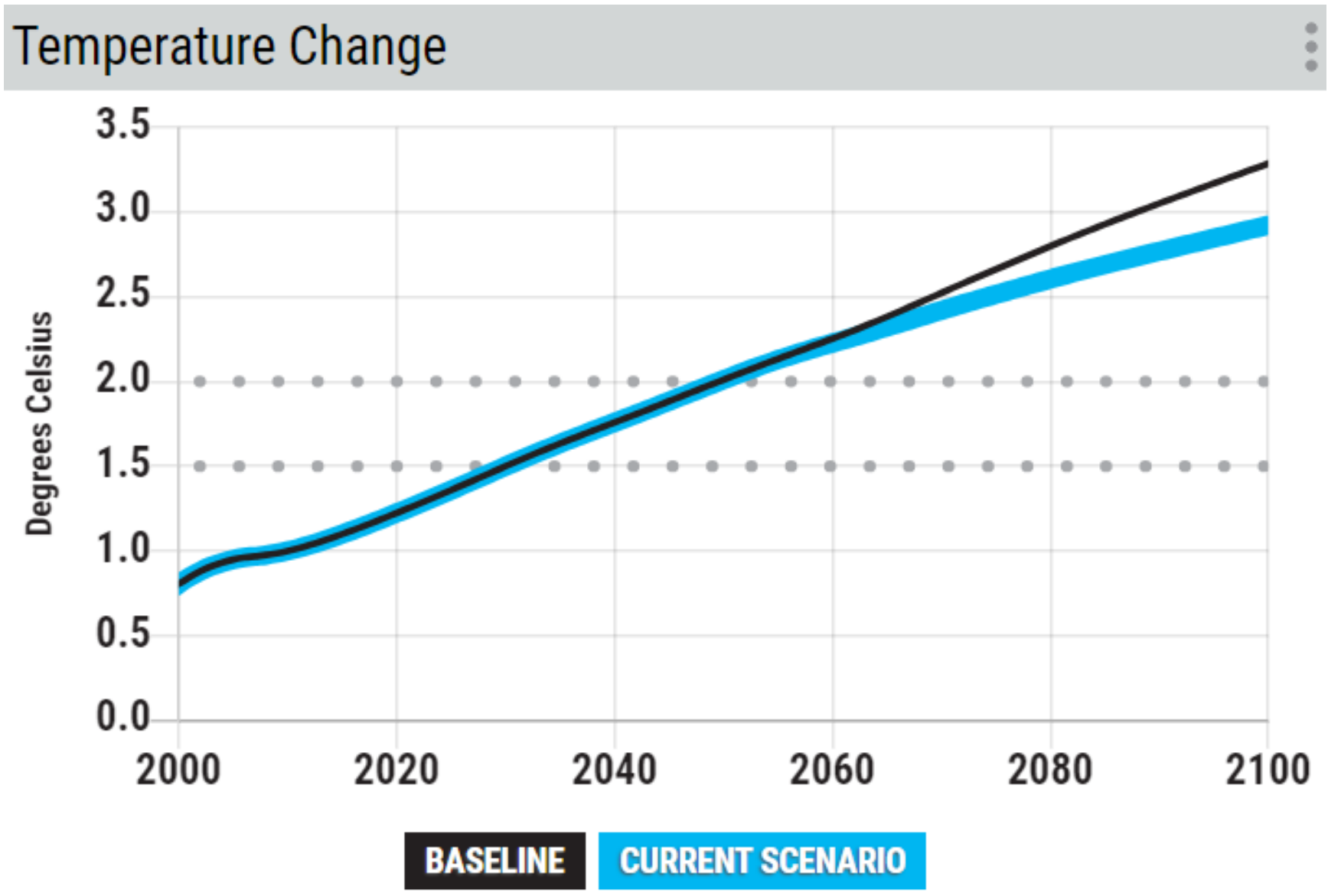

Figure 6: Temperature change with 10% US and 50% China emissions reduction starting 2050

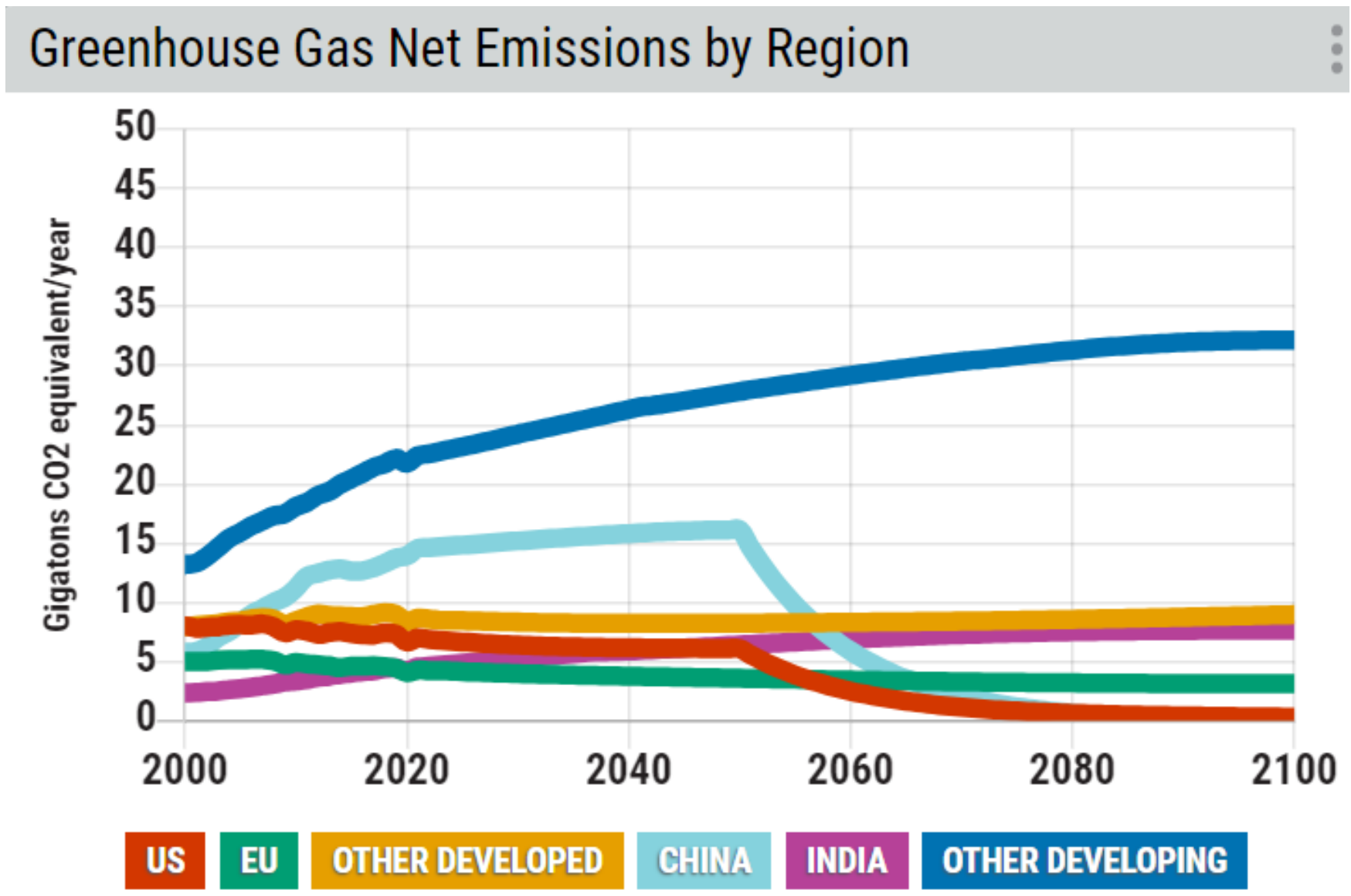

Figure 7: GHG net emissions by region with 10% US and 50% China emissions reduction starting in 2050

Considering other countries in the picture, a stark increase in temperature befalls the "other developing countries." Although strides must be made in an effort to decrease emissions in this region, due to the political instability, and infrastructure shortages, it may be beneficial to introduce green and sustainable infrastructure through other developed countries, who can provide more resources. As stated previously, focusing first on human health, and increasing life expectancy can cause these countries to increase their own output quickly, thus in turn creating capital to propel forward into green technology when communities are ready. In terms of the EU, India, and other developed countries, whilst making a change in the reductions of emissions would be a positive step forward, the most efficient impact can be created through analyzing the geopolitical and sustainable structure of the US and China. In Figure 7, it is noted that each of the remaining countries intersects China's emissions in the year 2057, at halfway. Given that the predicted model shows that these developed countries will be in the same place as China in 50 years down the road without any major sustainability efforts, it foretells the economical decision of focusing on the US and China to reduce emissions.

Assuming a world in which no country is able to feasibly conduct emissions reductions year after year, the introduction of a "peak emissions year" presents itself as a viable option. This policy option would entail each country pledging to reduce or maintain emissions after the year 2050, leaving more than 25 years to make changes and adjustments to meet future targets. Although this policy does not drastically reduce the projected global temperature increase (fielding a decrease of only 0.1° Celsius in global predicted temperature change), it sets up guidelines and processes for countries to follow through with reducing emissions in the future, without risking economic well being or trade relations.

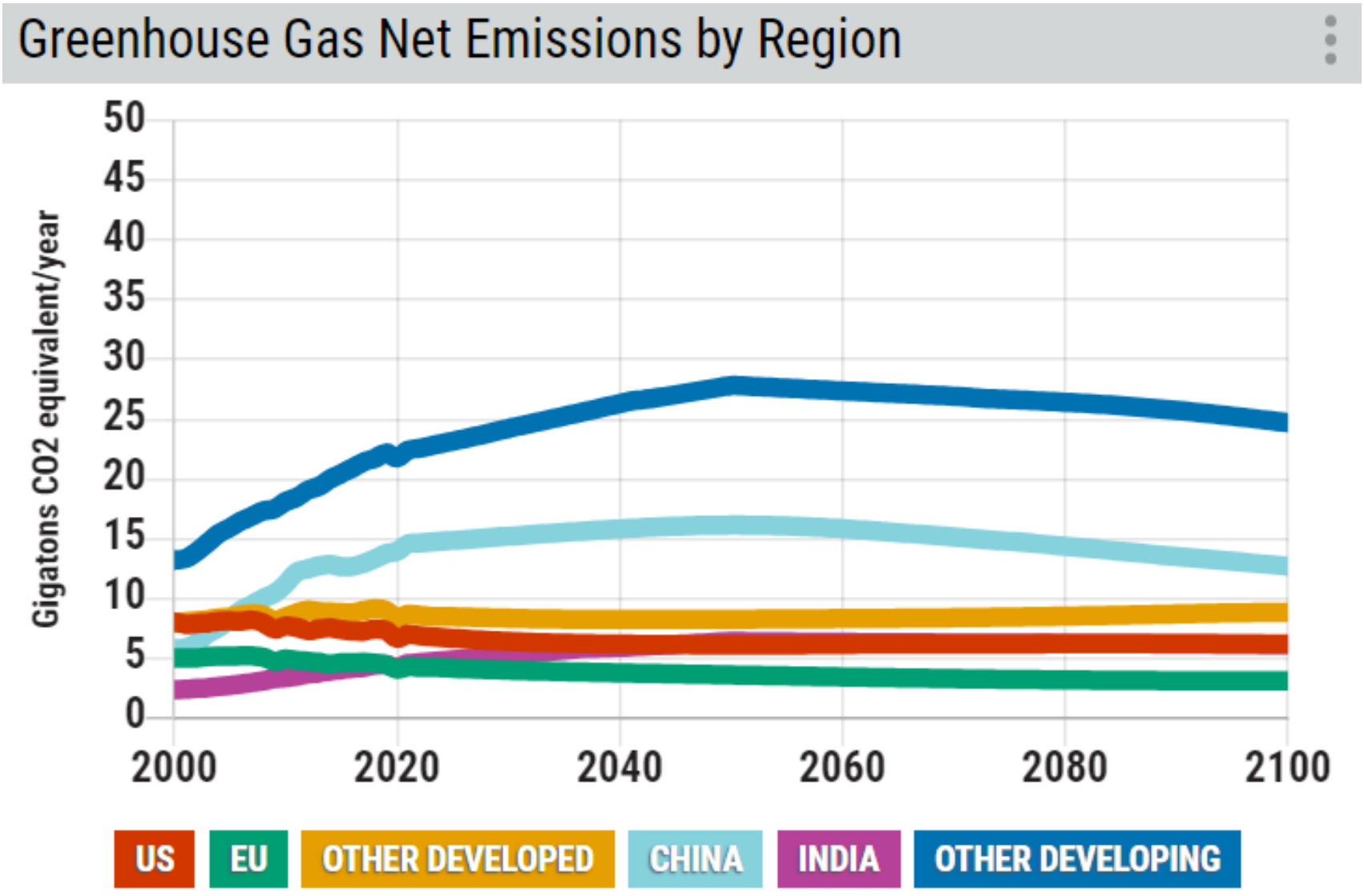

Figure 8: GHG net emissions with worldwide 2050 peak emissions

## 3.5 C-ROADS: Impact of $N_2O$ decrease

Although the environment is the first impacted by climate change, human health is inextricably tied to the environment, and atmospheric conditions. To analyze the impacts upon human health from emissions, a few other metrics were used to measure the impact on human health, primarily $N_2O$ content. For example, in China, by cutting back emissions by 10% annually starting in 2050, the $N_2O$ emissions stand to decrease tremendously, as evidenced in Figure 9. Reducing $N_2O$ emissions is imperative, as it has been evidenced to save lives. In a recent study from 2021, $N_2O$ emissions were attributed to higher COVID-19 and other respiratory disease cases (Anser et al., 2021). And, considering that expansionary monetary policy is positively correlated with $CO_2$ and other emissions (Liguo et. al, 2022), understanding the need to decrease emissions to save lives is imminent when enacting monetary policy.

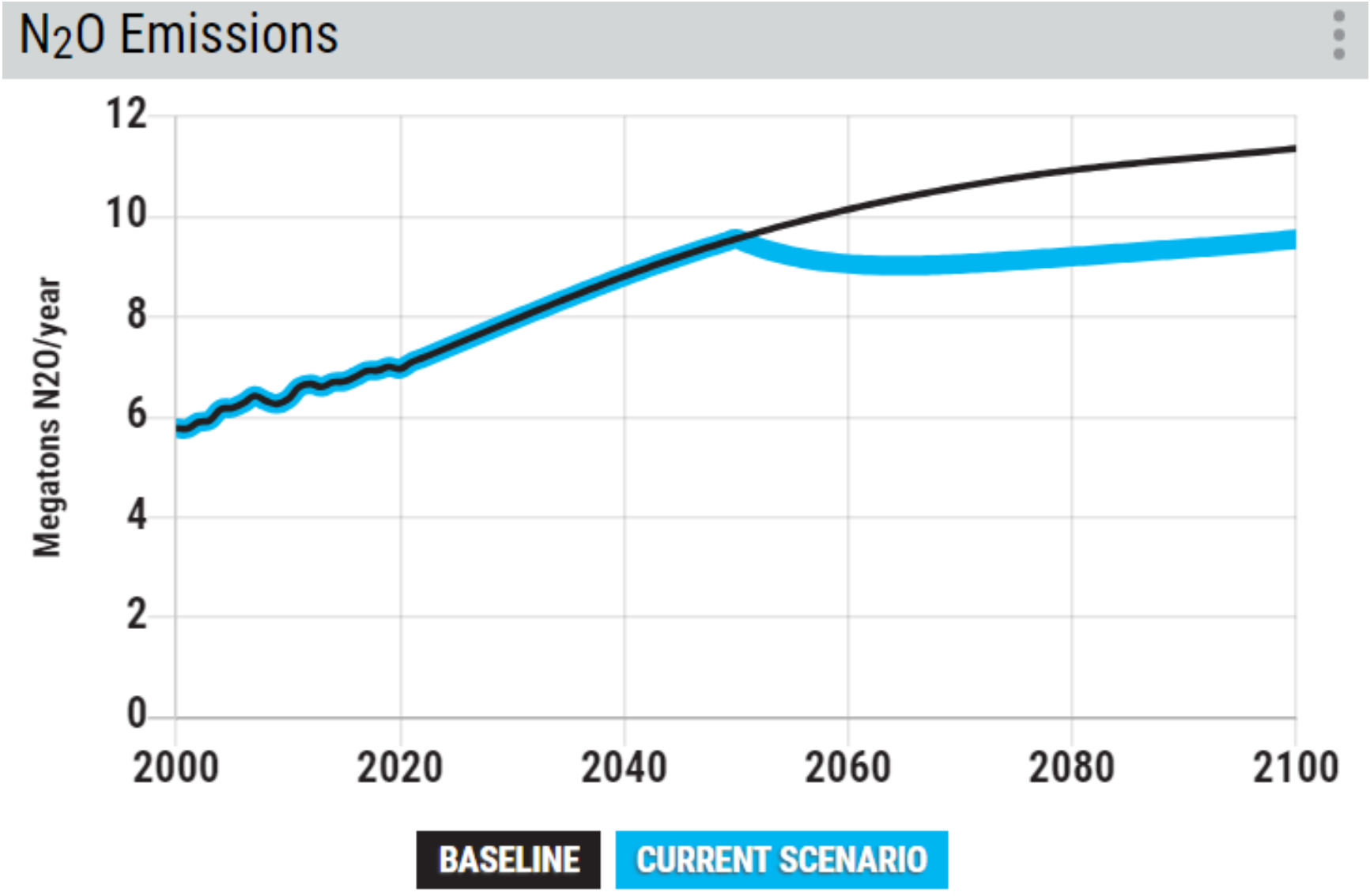

Figure 9: $N_2O$ emissions following China implementing a 50% decrease in emissions

Looking further into the Asian continent, by reducing emissions by 10% starting in 2050, India would also prevent massive amounts of $N_2O$ from being added to the atmosphere, amounting to around 2.7 Mts (see Appendix B). Saving these Mts from human inhalation could also alleviate burdens on the healthcare systems of both countries, providing a solution to decreasing costs and expenditures for government funded healthcare, and other costly programs.

In terms of other environmental benefits, this specific combination of reduction of emissions can offset a 0.4° Celsius increase in global temperature through the end of the century, further alleviating the positive feedback loop regarding the ice caps, and albedo of the Earth's surface.

## 3.6 C-ROADS: Impacts of Deforestation and Afforestation

As deforestation exhibits a very physical and evident example of human changes to the Earth's surface and climate, there are several other key factors of deforestation that go unnoticed, and must be uncovered. Deforestation can increase annual streamflow, and create a greater risk of flooding, due to the decrease in evapotranspiration from previous vegetation (Zhang et al., 2017). Coupled with greater risks of loose soil due to a lack of roots from vegetation to keep the sediment in place, this can also create hazards for rock sliding, and other seasonal natural disasters. While this may seem negative by itself, another method uncovered to remediate this has disrupted ecosystems. In areas where stream flow must be decreased, planting fast growing exotic plants have become a popular option. However, these exotic plants can disrupt the natural ecosystem (Filoso et al., 2017) adding to the burdens linked to deforestation. Although the initial root cause may not seem serious, dangerous fixes akin to the aforementioned can wreak havoc on an ecosystem. Along with this, forests are also important due to their health benefits and impact on the atmosphere. Forests and vegetation can add back necessary atmospheric moisture to the air, which can cause an increase in precipitation downward (Creed et al., 2018). This may be very helpful in areas stricken by droughts, as low maintenance plants can ramp up the process of greening. Another key idea policymakers can implement is the strategic position of forests adjacent to farmland. This can increase the crop yield due to increased water, yet in a natural method, which does not upset the ecosystem. As afforestation becomes a tactic that is increasingly widespread and necessary for survival, small considerations for the location of forests can have a positive impact on the economy as well.

In recent years, deforestation has been caused by several factors that change the location and frequency of deforestation. These key factors include agriculture, urbanization, and logging (Zhang et al., 2021). In fact, industry is a large contributor to the increase of deforestation in recent years, even for plants that may not seem connected to the deforestation business. For example, 20% of deforestation in the Michoacán region comes from a US demand for avocados (Cho et al., 2021). Whilst these seemingly harmless crops may not appear to be the driving force behind deforestation, they can be. In addition, local governments can also be drivers behind deforestation, as seen in Brazil in regards to the Amazon Rainforest (Peres et al., 2023).

However, as many governments are looking towards afforestation as a solution, it is essential to consider the specifics of where and what to plant. Whilst some may object that afforestation and population growth are not compatible, historically, in New England, there was an increase in both forestation and people onwards from the 1850s, as they paid attention to responsible land use (Pfaff, 2000). Along with the increase in life expectancy and output afforded by the increase in greenery, green technology can make urban sprawl less common, creating communities geared towards efficient and comfortable solutions. An interesting finding by a study done on retrospective deforestation indicated that in 1000 BC, there was an extensive bout of European deforestation (Kaplan et al., 2009). The fact that Europe can come back to lush green forestry in the manner it has today, gives hope for future afforestation efforts.

In the status quo, in India, the current estimate of gross deforestation is low (-0.43%), indicating movement toward the right direction. However, there is regionality to this statement, as some regions have virtually no deforestation, whilst others, such as the Deccan Peninsula face rates of up to -3.2% (Reddy et al., 2013). This regionality in deforestation can also be seen in China, as there are large differences in the locations of deforestation in China (Zhang et al., 2022). However, there have been positive strides towards afforestation, as in 2018, the Chinese government reaffirmed plans to increase national forest coverage to 23% (Stunkel and Tucker, 2020). China has also seen rapid growth in afforestation over the past few years, as China's forest cover has grown from 8.6% in 1949 to 18.21% in 2003 (Zhang et al., 2006). Taking note of the positive direction of change, in these simulations, China was placed as the forefront player in afforestation, reaching a maximum of 50%, alongside the US and EU at 10%, as they gain traction in afforestation techniques. A key impact of deforestation is the decrease in carbon in the atmosphere, and as evidenced by Figure 10, the gross $CO_2$ removal from afforestation will be almost 2 Gtons of $CO_2$ per year. Considering that increased trade between the US and China is affecting the forest reserve of the Amazon basin and the Sub-Saharan desert, known as the lungs of earth (Kumar, 2022), the net $CO_2$ removed is lower, at approximately 1 Gton/year.

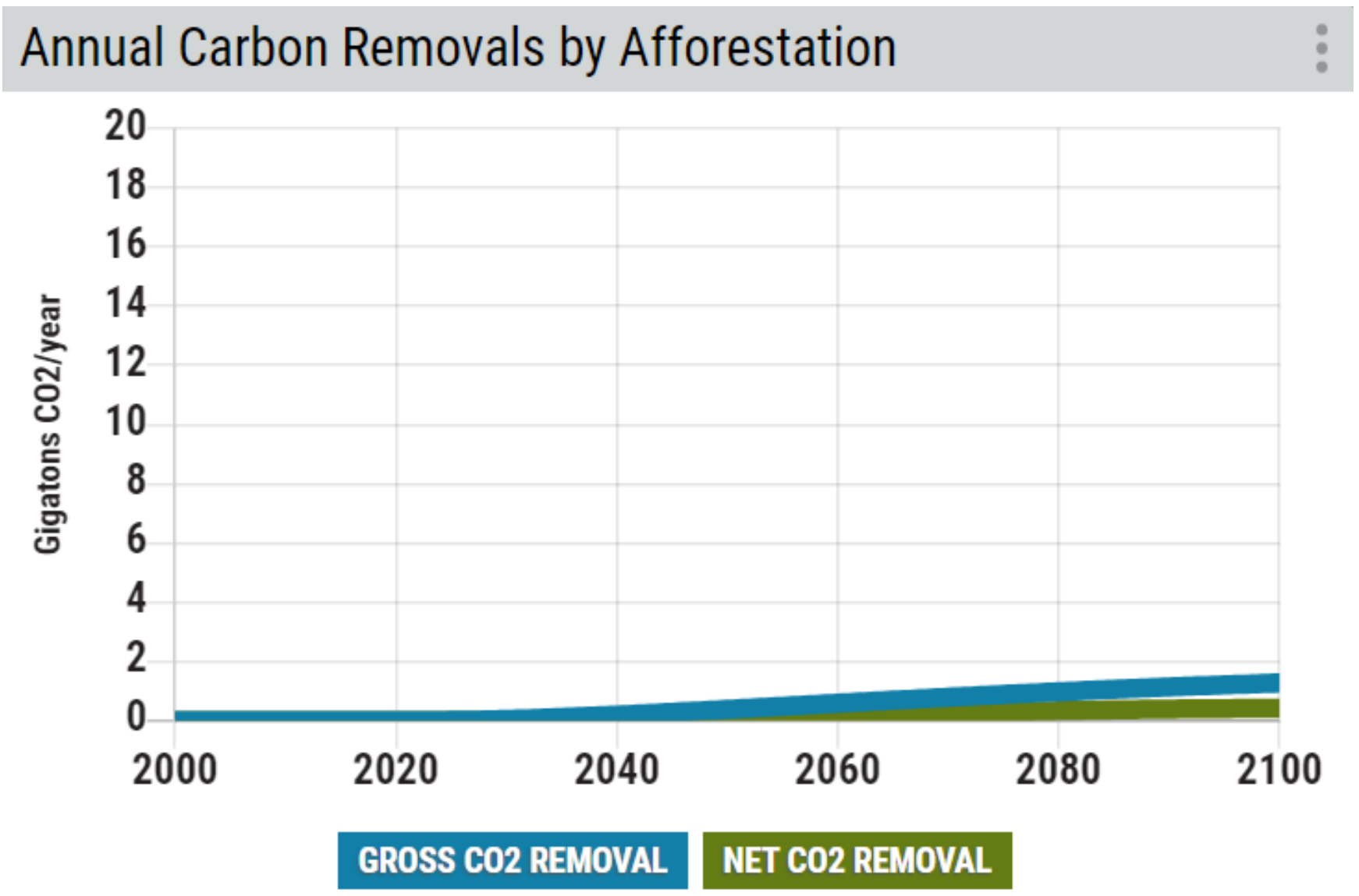

Figure 10: Annual carbon removals with 50% China, 10% US and EU afforestation

As the nation-group of 'other developing countries' has been grouped together, it has been found that a significant push for preventing deforestation in these countries (upwards of 50%) can eliminate up to 112 Gtons of $CO_2$ by the end of the century, or approximately 1.12 Gtons/year of removed $CO_2$. Additionally, a previous study by Wyatt et al. (2022), considered a 5.1% yearly decrease in deforestation per year globally, which resulted in a decrease in average global temperature change to only 2° Celsius, indicating the positive effects of preventing deforestation.

The policy impacts of preventing deforestation are numerous. In tandem with the environmental impacts, increased air quality presents another added benefit of leading to fewer respiratory diseases, which can lessen the extent of the burden of healthcare systems on government policy in developing countries.

## 3.7 En-ROADS: Status Quo

In Figure 11, the auburn, brown, and blue bands represent the largest shares of global energy production today: oil, coal, and gas respectively. Policy initiatives have been targeting these sources of energy recently, trying to shift consumption towards more renewable sources of energy, and away from these fossil fuels. Interestingly however, it is clear that increased work needs to be done to address the difference between the energy share of renewables, bioenergy, and nuclear energy, in comparison to oil, coal, and gas. This places a greater emphasis on the development of technology to increase the share of energy produced by bioenergy, renewables, and nuclear energy.

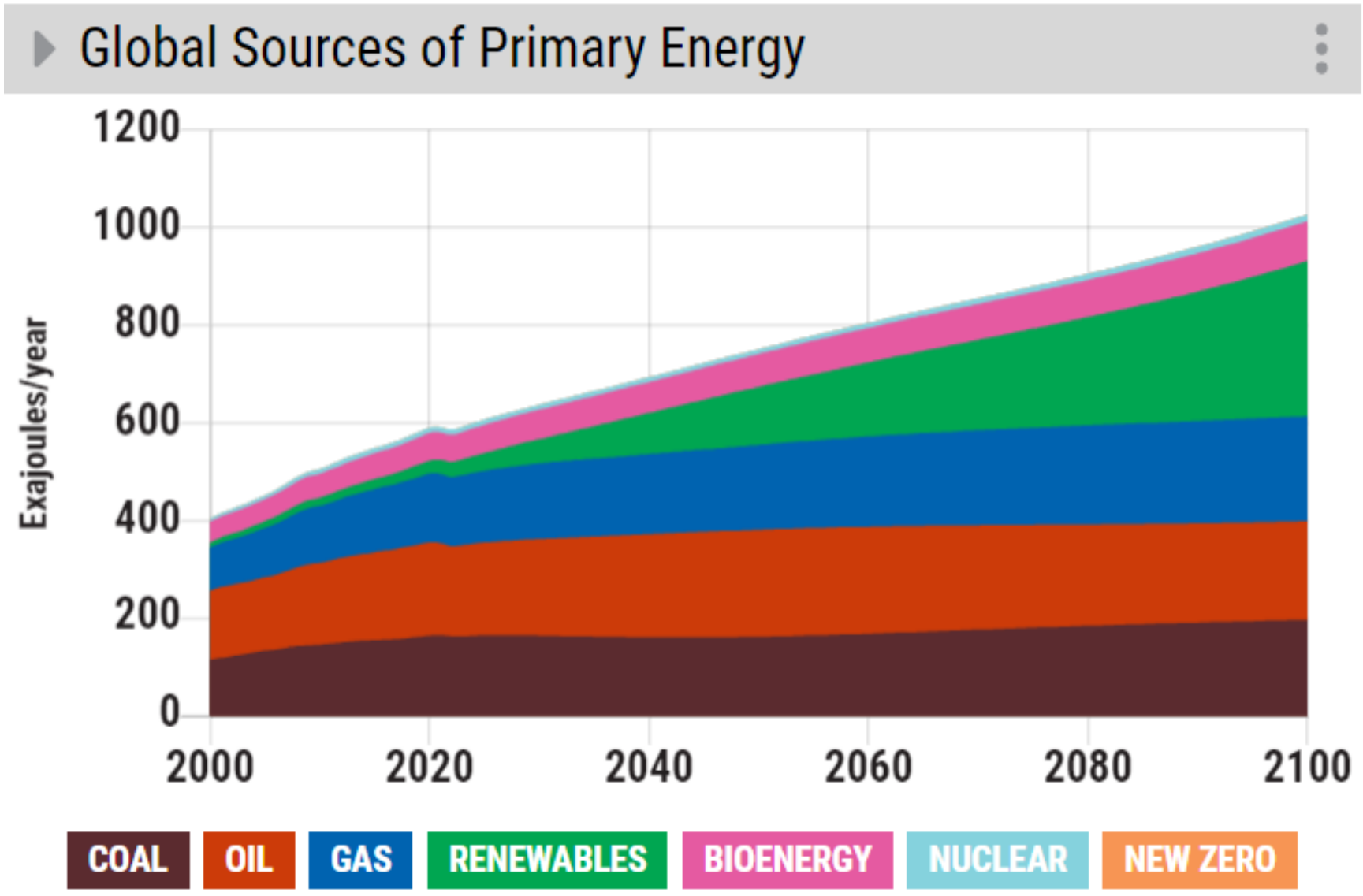

Figure 11: Global sources of primary energy in the status quo

## 3.8 En-ROADS: Impacts of Taxation & Prices

Taxes have frequently been used as a method by which to alter consumer behavior and deter further purchases. When analyzing which tax would be most effective to eliminate the reliance on non-renewable energy sources, putting a price on carbon emissions seems to decrease the share of non-renewable energy sources significantly. In fact, global carbon pricing revenue has increased over 60% to $84 billion from 2020 to 2021. Additionally, there are over 68 different operating carbon pricing instruments today, making this emission price feasible to fix (World Bank, 2022). By increasing the price of carbon emissions per ton to $40 for firms (as seen in Figure 12), this greatly reduces the share of nonrenewable energy production, as renewable energy increases significantly. Although there is a small dip in energy production, towards the end of the century, the amount of energy production is quite similar, showing a resilient global market which rebounds from taxation. To compare this more moderate taxation approach with a tactic more extreme, a side by side analysis with a $250 tax per carbon was also set in place. From the primary sources of energy, it becomes apparent that renewables took on most of the energy burden, however the overall energy in exajoules per year still decreased, indicating that there would be a shortage, and misallocation of resources. An interesting observation is the fact that bioenergy and other types of energy do not play into the market as much as renewables do, meaning that a small push for more renewable energy may incur big returns.

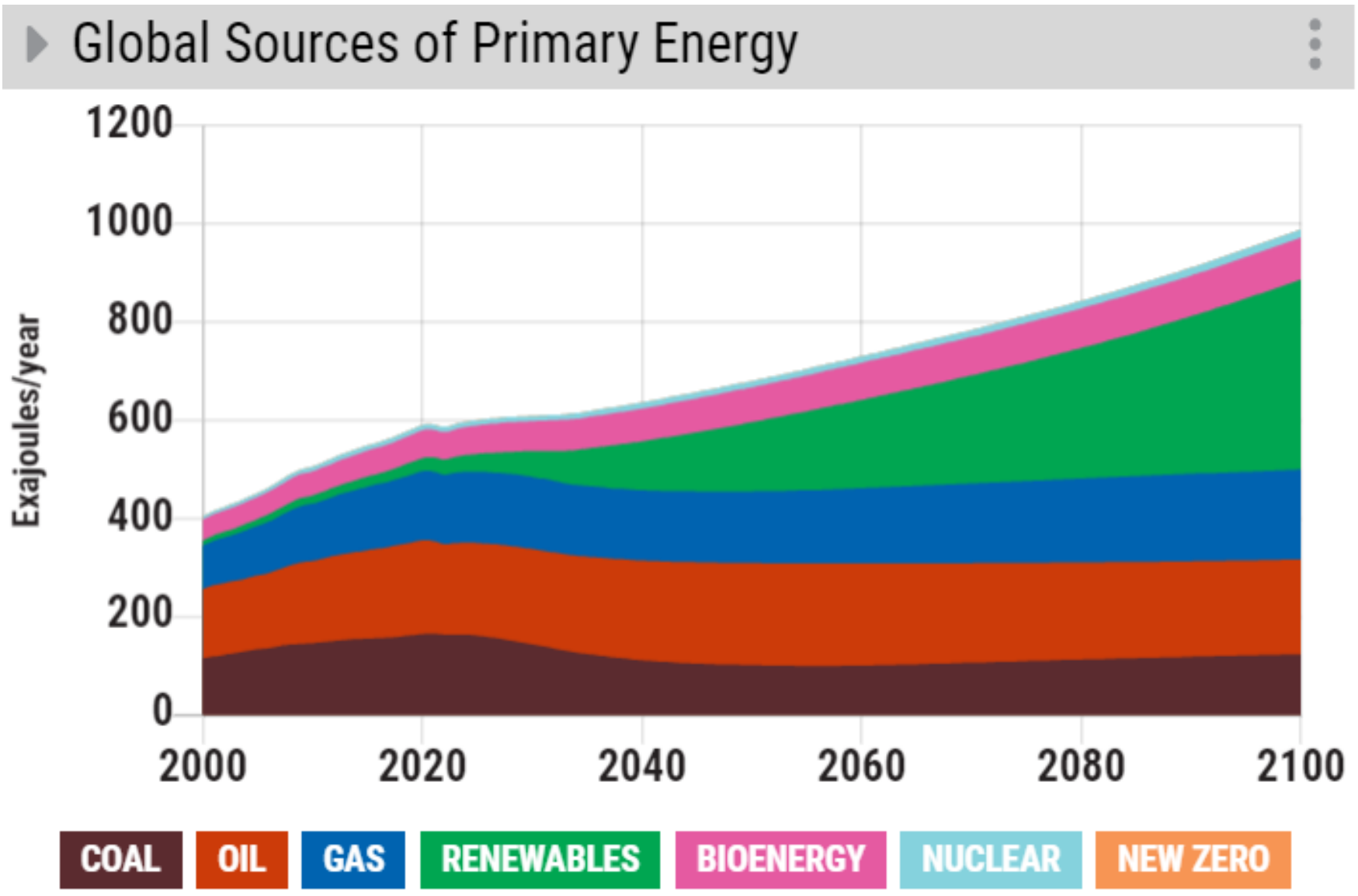

Figure 12: Global sources of primary energy with a $40 carbon tax per ton emitted

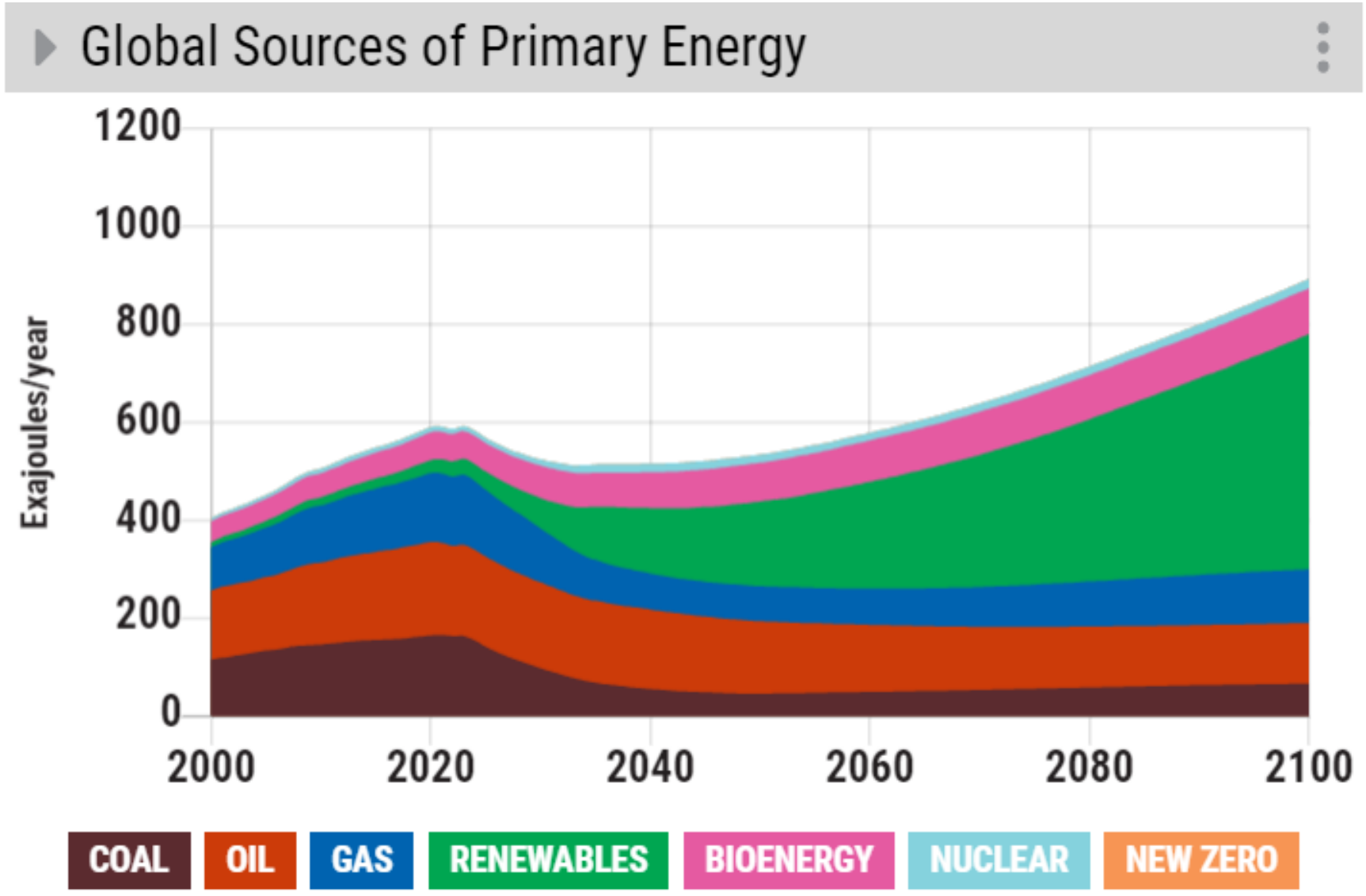

Figure 13: Global sources of primary energy with a $250 carbon tax per ton emitted

Looking further at the marginal cost of electricity production, increasing the tax on carbon to $40 causes minor fluctuations in initial years on the marginal cost of electricity production. However, taxing $250 for carbon causes a sharp rise in the marginal cost of electricity production. It becomes interesting to note the patterns after this sharp uptick, including the sustained high marginal cost for oil, and a subsequent decrease in the marginal cost to their original levels for coal and gas.

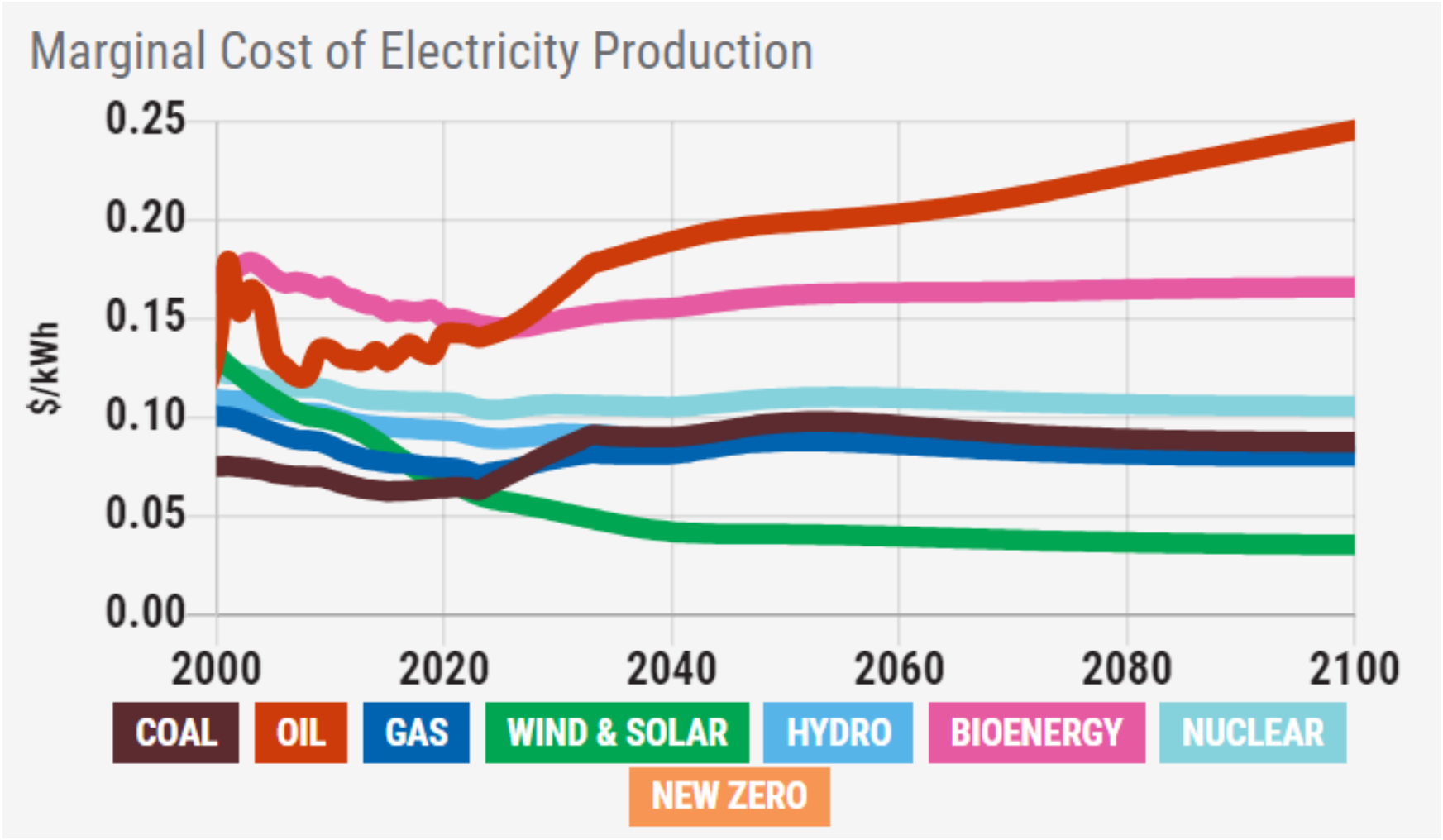

Figure 14: Marginal cost of electricity production with a $40 carbon tax per ton emitted

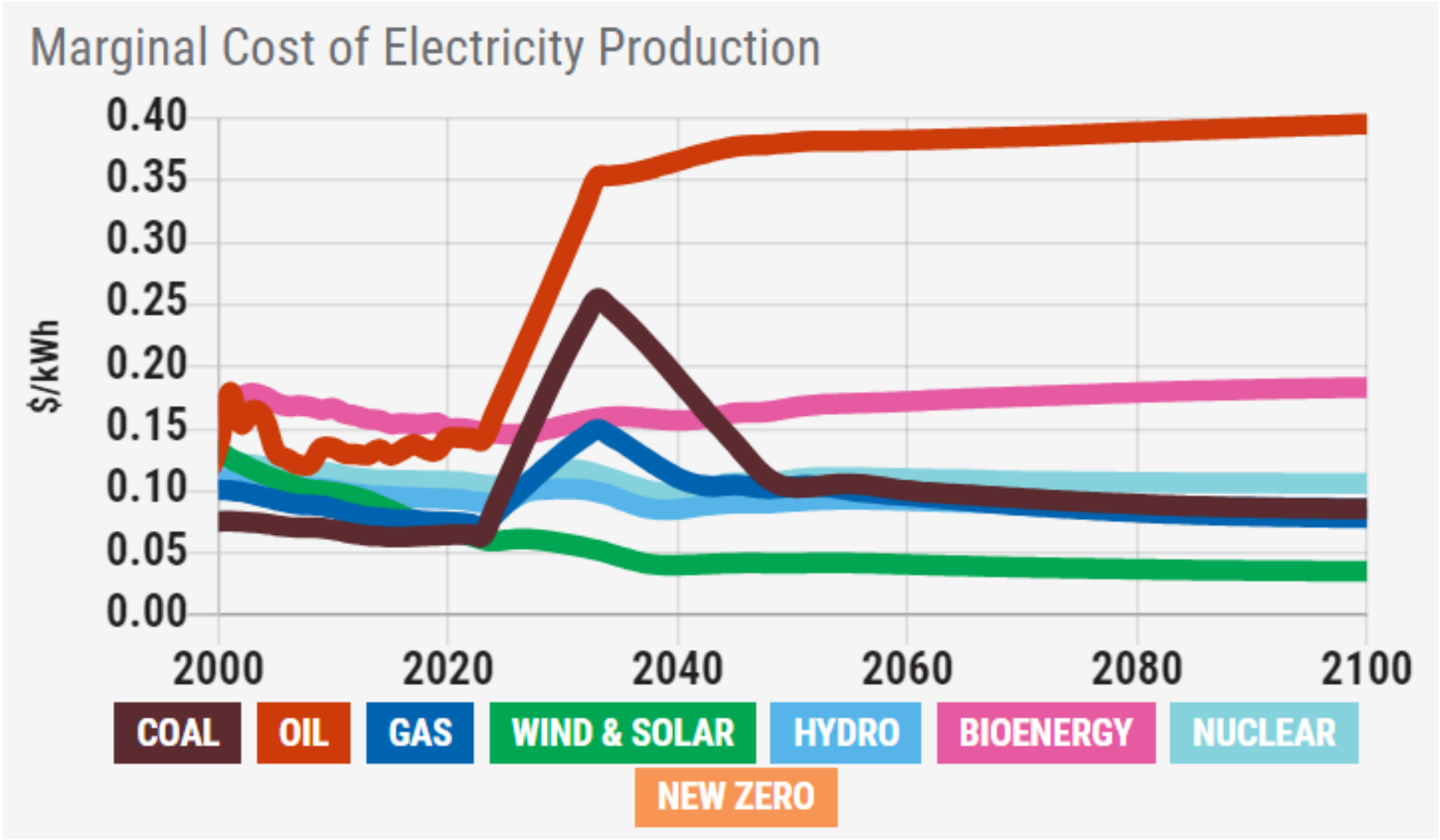

Figure 15: Marginal cost of electricity production with a $250 carbon tax per ton emitted

Thus, increasing the price of carbon emissions per ton to $250 increases the share of renewable energy, but also depletes the total energy created, which could have other disastrous effects. Moderately increasing the price of carbon dioxide emissions to $40 per ton creates a balance between regulation and efficiency, without sharp upticks in marginal cost that would create supply chain disruptions.

Further looking at the impacts of a $250 tax on carbon, the ice albedo effect also becomes a factor. This arctic amplification of climate change occurs because of differences in surface albedo feedback, and longwave feedback, which can come back and trap large amounts of heat (Winton, 2006). After introducing a prohibitive tax on carbon emissions, it becomes clear that this is one of the most direct ways to combat global warming, due to its impact on the ice sheets. An "ice-free" arctic summer is defined as a

September in which the total area of the Arctic ice sheet is less than 1 million square kilometers, or less than 7% of the Arctic ocean area (Sigmond et al., 2018). The probability of an ice-free arctic summer in 2100 is almost reduced in half, along with a 7% reduction of an ice-free Arctic summer in 2050 with a steep carbon tax introduced. Although the probability of at least one ice-free Arctic summer before 2100 is still 100%, the most impactful takeaway is that the direction of change towards slower melting rates is positive.

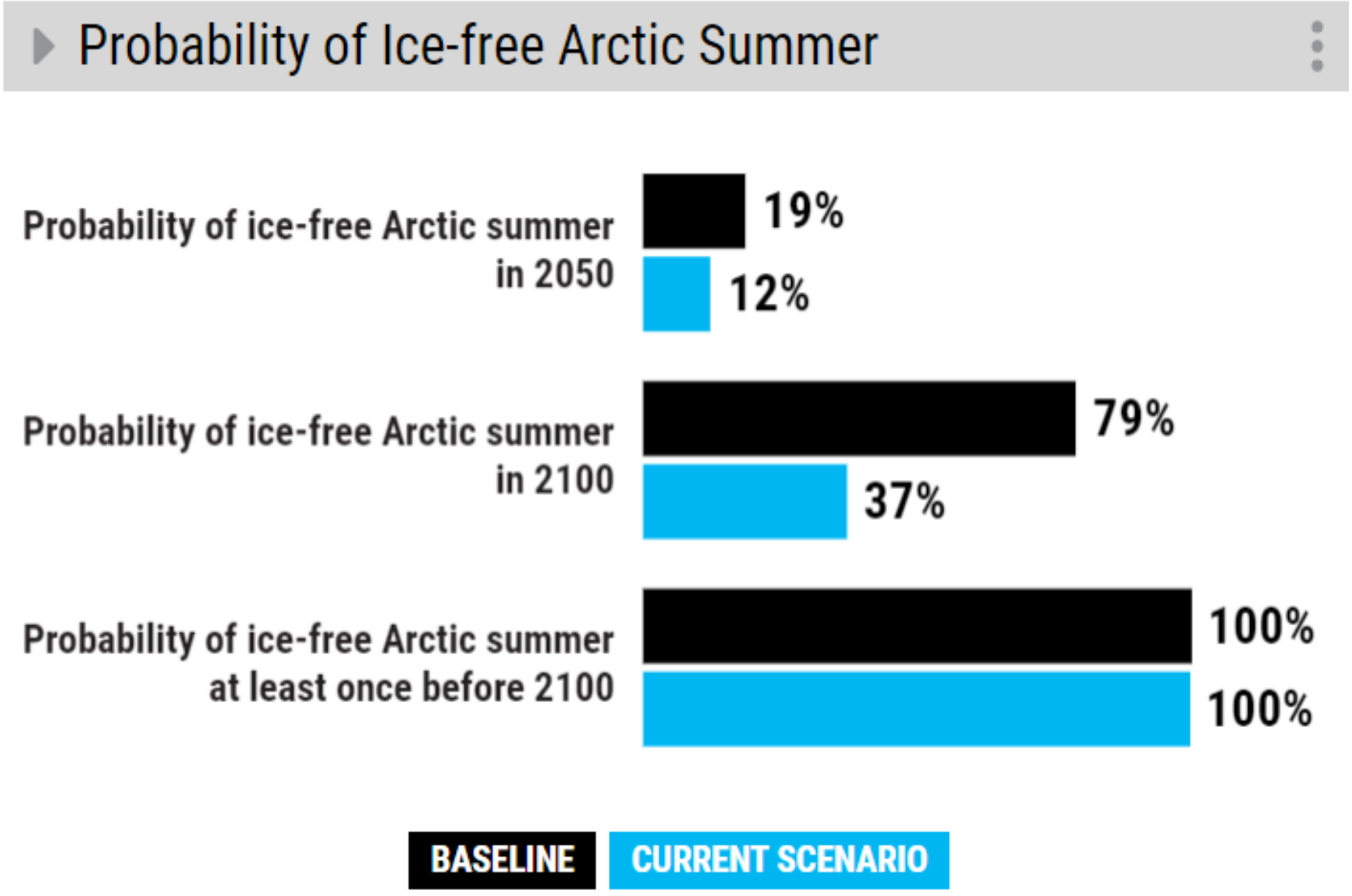

Figure 16: Probability of ice-free arctic summers with a $250 carbon tax per ton emitted

Another sector that could be taxed is the coal industry, and the results show it may be better to incur reductions in coal utilization earlier rather than later. Assuming an eventual 33% decrease in the usage of coal through the implementation of other renewable technology, decreasing production sooner proves beneficial, as when comparing primary energy demand for coal, there is a less steep drop experienced if reduction is implemented in the next decade, rather than three. This results from the ever-increasing demand for coal as a primary energy source, and so to mitigate loss, it may be best to reduce coal usage gradually, but over a shorter time frame. The analysis for the 2030 gradual reduction of coal also presents an opportunity to reduce reliance on coal more than a reduction in 2050, creating an overall net benefit. Past studies regarding coal taxation on En-ROADS taxed coal starting in 2030, and saw an overall reduction of 25.34 gigatons of $CO_2$/year by 2100, given a 5 % increase per year (McNicholas, 2023). As this study noted the importance of reducing coal specifically due to its non unique output to the energy sector, beginning coal taxation early on in the climate change mitigation process proves to be a viable option.

As the driver behind a large portion of the transportation industry, and thus incurring other spillover effects, the price of oil can change the outlook of many different industries. Its contribution to the current energy production distribution is significant, as by decreasing the quantity of oil produced by means of a tax, total energy production also increases, indicating that there is no substitution effect of increased renewable energy, or further innovation.

As oil is taxed to a prohibitive level however ($100 per barrel of oil), it becomes evident that innovation and increased reliance on renewable energy creates a more likely scenario than a complete shift to other fossil fuels, such as coal. The greatly increased proportion of energy production in an extreme scenario represents that advancement for renewable energy is propelled by need, and not proactive innovation.

## 3.9 En-ROADS: Impacts of Building Innovation & Retrofitting

As planning and urban development come to the forefront as ways to mitigate the impact of excessive energy use, it becomes imperative to model new buildings, and their energy efficiency. Residential and commercial buildings are important targets, as they represent 20% of the total global energy consumption (De Paola et al., 2014). Recent trends have seen energy efficiency for households increase, but decrease subsequently as the homes and buildings become larger (Desilver, 2015). Given the scenario that new buildings become 3% more energy efficient each year, with approximately 8% of older buildings retrofitted per year, the global predicted temperature increase by the end of the century is approximately 3° Celsius, or .3° Celsius less than the status quo. In a previous study on retrofitting and building innovation through En-ROADS, with a marginal increase of 3.5% energy efficiency increases each year, a large spike in the price of energy leading up to the year 2040 was observed (McNicholas, 2023). In this approach, a lower energy efficiency increase was chosen, to mitigate the adverse economic impacts and supply chain blockages that may arise from an increase in energy prices.

Another interesting aspect of this push for energy efficiency in newer and older buildings alike is the shift in the GDP away from energy revenue. As seen in Figure 17, the amount of worldwide Exajoules produced per GDP in trillions decreases, demonstrating other examples of industries that may pick up in place of energy, and the shift away from energy to power increasing GDPs. As countries move further and make progress on their energy policies, it seems as though collective action amongst a block of countries may bring about more change than individual countries themselves. In West-Africa specifically, it was found that regionally coordinated policy was effective, especially when following the guidelines of international goals, such as the UN Sustainable Development Goals (Antwi-Agyei, 2018). This becomes important to note as developing countries look for ways to influence their industries and become more sustainable.

In an ideal world, with a higher rate of energy efficient housing production, of around 5%, the energy landscape seems to shift away from coal, oil, and gas, and increase dependency on bioenergy and renewables. This represents the most direct way to limit the production of fossil fuels and other nonrenewable energy sources, cutting out the demand for these types of energy. While in total, the final amount of energy produced is much less over the course of the century, in the long run this is feasible because of the smaller collective demand for energy to begin with.

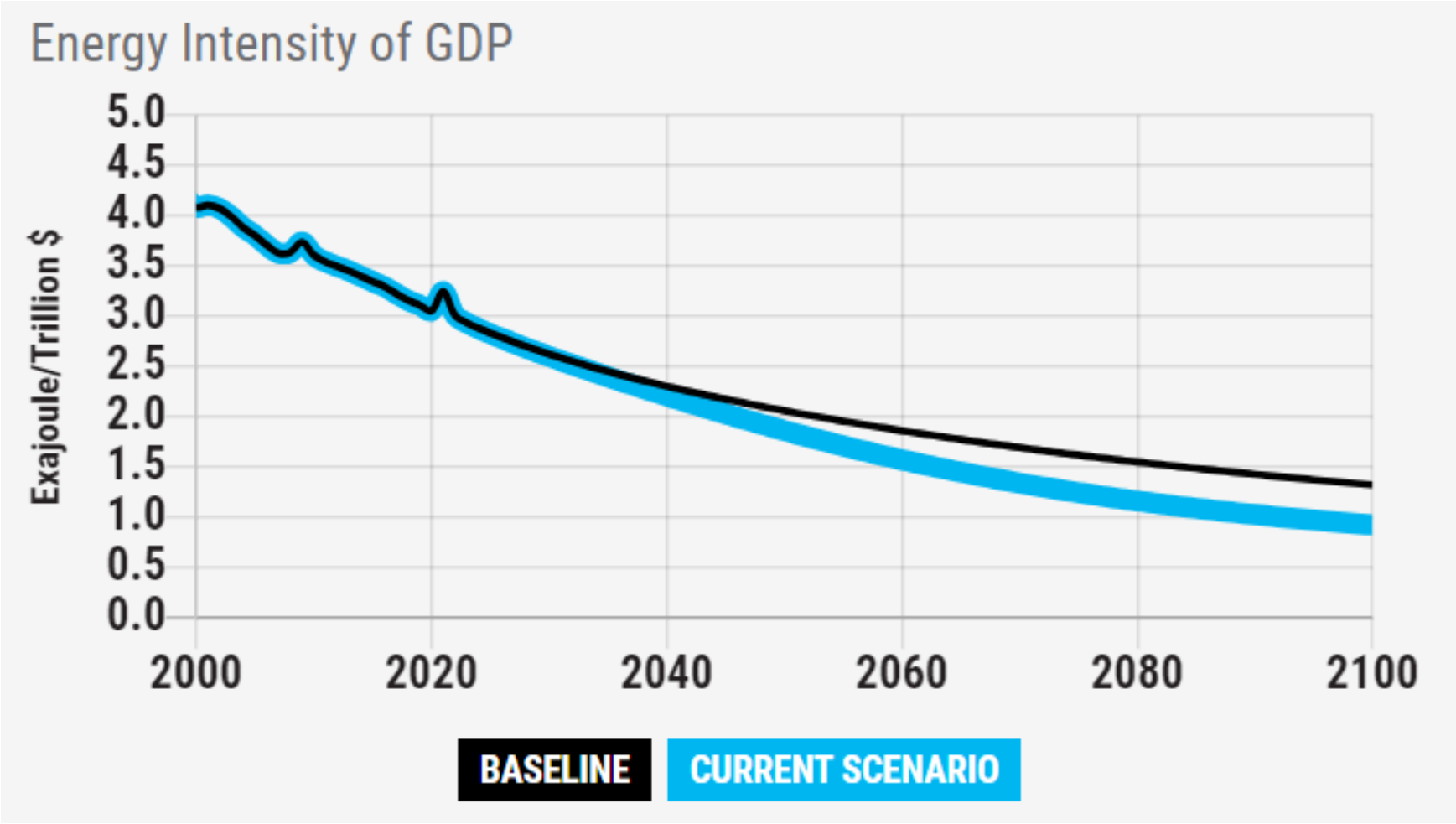

Figure 17: Energy intensity of the GDP with a 3% energy efficiency rate, and 8% retrofit rate

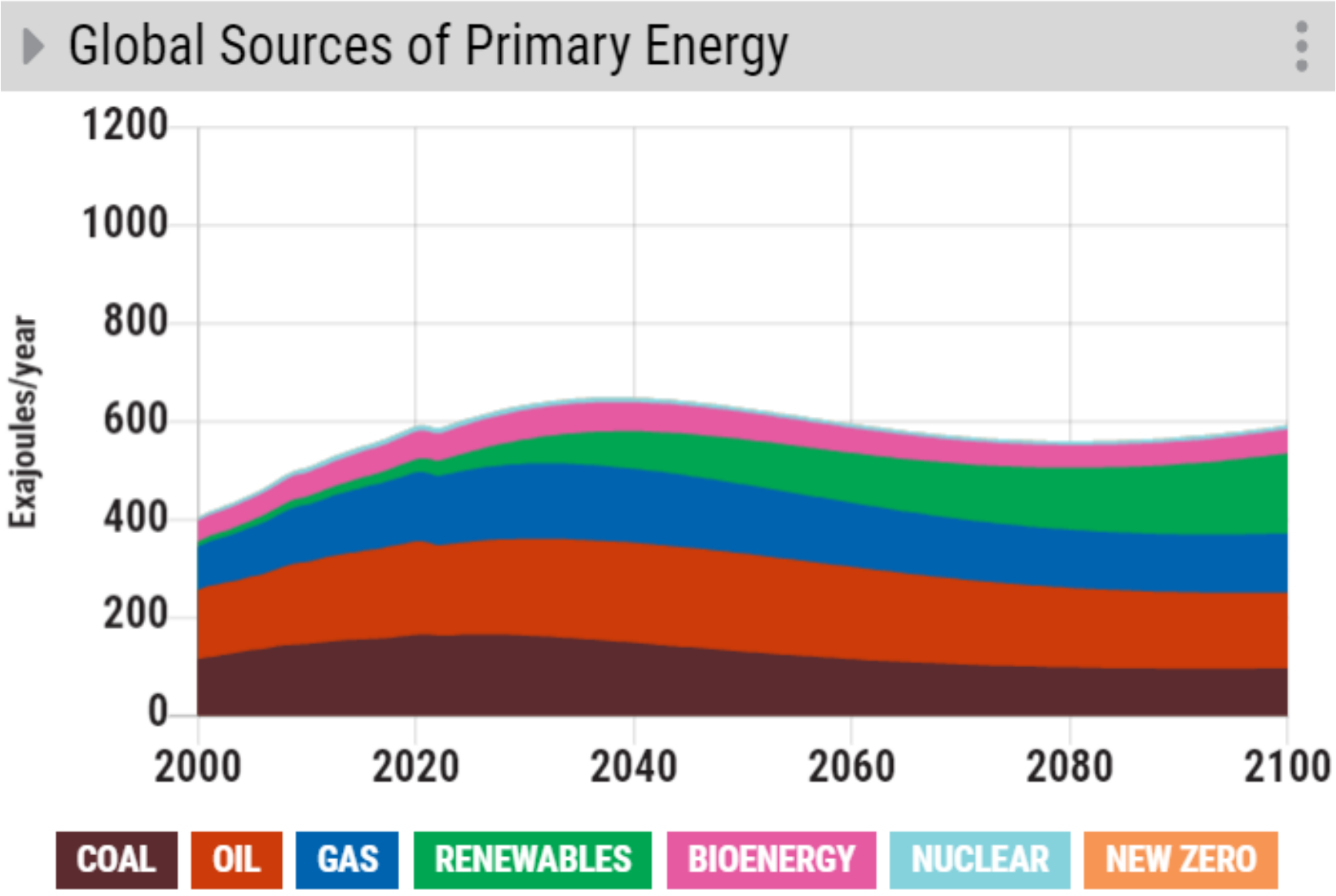

Figure 18: Global sources of primary energy with a 5% energy efficiency rate

## 3.10 En-ROADS: Impacts of Innovation & Subsidies

As technology is increasing the quality of life for many across the globe, it is also proving an important tool to fight against climate change. Through subsidies, the government can incentivize firms to produce more bioenergy, and other forms of renewable fuel. A moderate subsidy for bioenergy of $10 per BOE can increase the primary energy demand for bioenergy by over 11%, thus increasing the share of eco-friendly fuel in the market. A more drastic subsidy of $30 per BOE results in an over 37% increase in the primary energy demand for bioenergy, representing the positive potential existing for bio-based energy sources.

Subsidizing renewable energy per kWh can also increase the primary energy demand for more environmentally friendly energy sources. By subsidizing $0.02 per kWh, the primary energy demand for renewables increases by more than 30%. A slightly larger subsidy of $0.03 per kWh can increase primary energy demand for renewables by upwards of 50%. A comparable study on En-ROADS introducing a subsidy of $0.01 per kWh also sees the marginal cost of renewable energy (namely wind and solar) decrease by more than half (McNicholas, 2023), indicating the positive effect of renewable energy subsidies. As prices for a kWh of energy are typically $0.10, a marginal increase in price for a relatively large increase in the demand for renewable energy poses moderate renewable energy subsidies as a wise choice for decreased reliance on fossil fuel energy.

## 3.11 En-ROADS: Impact of Policy Timeframe

Policy initiatives posed by many governments have varying due dates and end goal years that typically range from 10-60 years into the future. Although sooner change may be the most appealing for immediate impact, upon examination of different lengths of policy initiatives, the efficacy and impact overtime of these measures is highly varied. A shorter time frame may provide a shorter reduction in methane production (10 years), but after the 10 year threshold has been hit, methane production skyrockets once more, indicating that the solutions posed were never permanent. For the 60 year production time frame, although the decrease in rate of change of methane production is slower, the growth is slowed consistently for all generations, ensuring that certain age groups are not disparately impacted by a surge in the production of methane after a strict policy reduction.

On the other hand, the production of oil and new zero carbon breakthroughs were also examined to determine the effect of policy timelines on efficacy. Assuming a high tax on oil, of $85 per BOE, if this tax is imposed in 2060, approximately 35 years into the future, the global temperature increase will be the same as the status quo over the next century, of 3.3° Celsius. In fact, the year-by-year temperature change graph is remarkably similar to the baseline, without any large tax changes. This is important to note, as an $85 tax on oil per barrel is significant, and would cause supply chain shortages, however even after all of the negative impacts the overall benefit to the environment would be the same as taxing a much smaller amount, starting in the next five years. Another example of this concept would be with new zero carbon. Assuming a slight breakthrough in this type of energy production, but only in 2060, with 7 years for this type of technology to become mass market from its initial price of double that of coal, the total amount of energy in Exajoules/year is strikingly similar to the status quo, even after a large breakthrough. Similarly, in a study done on China, it was shown that gradual implementation of policies that start in the near future prove the most effective for reducing reliance on nonrenewable energy (Debelle, 2019). These results show that even with large policy changes or discoveries, if they occur too far into the future, the final environmental results are diminished, and thus gradual change must be enacted, starting in the nearer future.

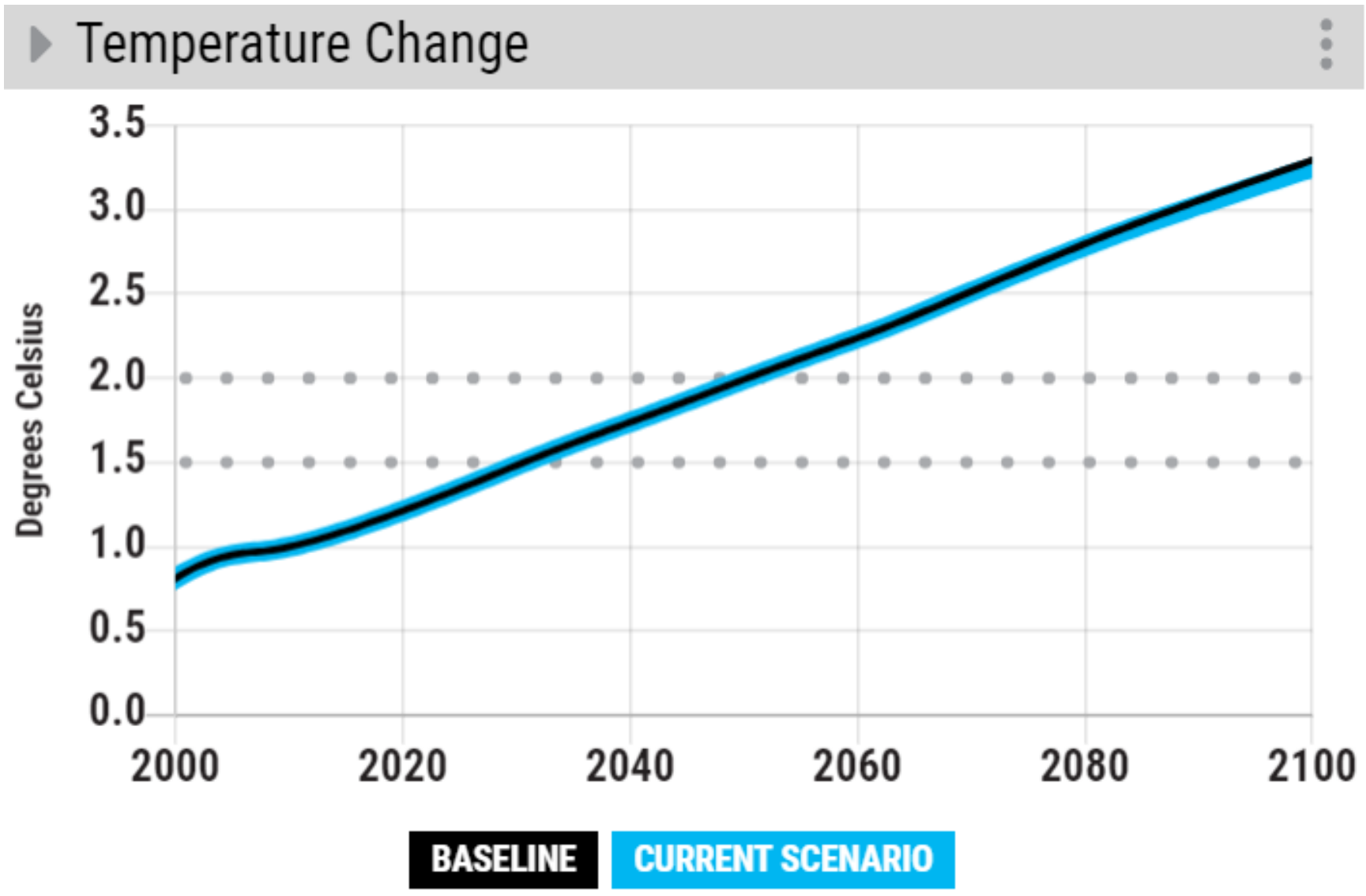

Figure 19: Temperature change with a tax of $85 per barrel of oil energy enacted in 2060

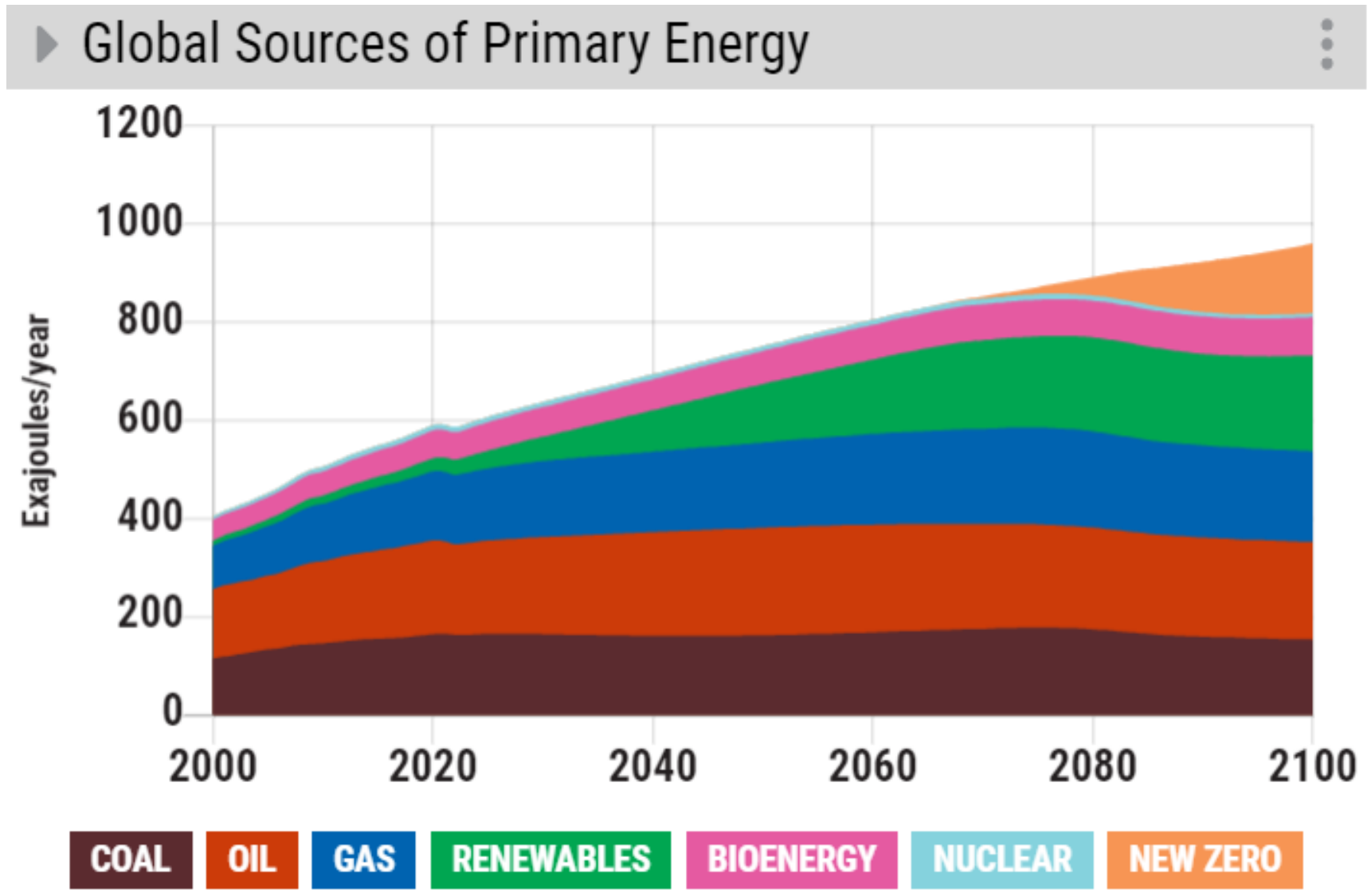

Figure 20: Global sources of primary energy given a new zero carbon energy breakthrough with 7 years of introduction at 2x coal pricing in 2060

## 3.12 En-ROADS: Impacts within Economic Markets

When analyzing markets, it is imperative to consider the supply and demand curves in conjunction with the price of each type of energy. Energy pricing is a vital component of the economy, as its prices impact the manufacturing of almost every consumer and capital good on the market. To analyze the impact of the gradual introduction of taxes on the prices of electricity, the price of carbon emissions was changed once more. Introducing a $250 tax on carbon, creates a very high peak and low valley in the market price of electricity. This is due to the loss of demand as the price for carbon increases. If this tax were to be subtly built in over the timeframe of 30 years, the peak and trough of the market price graph is much closer to

the baseline market price. This may be good, as it influences consumer decisions to stay away from nonrenewable sources of energy, whilst also avoiding extreme disruption to the market price of electricity.

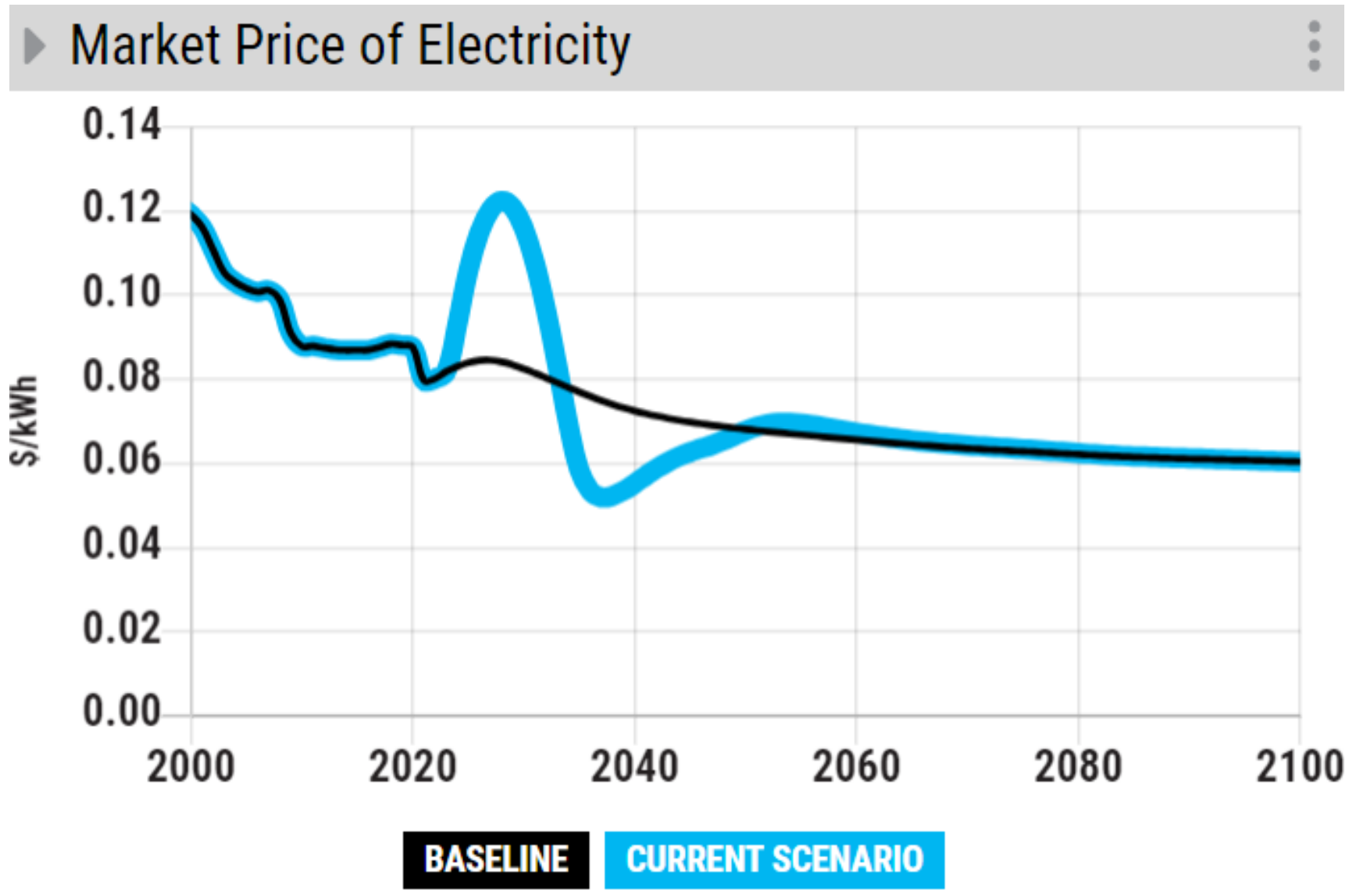

Figure 21: Market price of electricity with an immediate $250 carbon tax per ton emitted

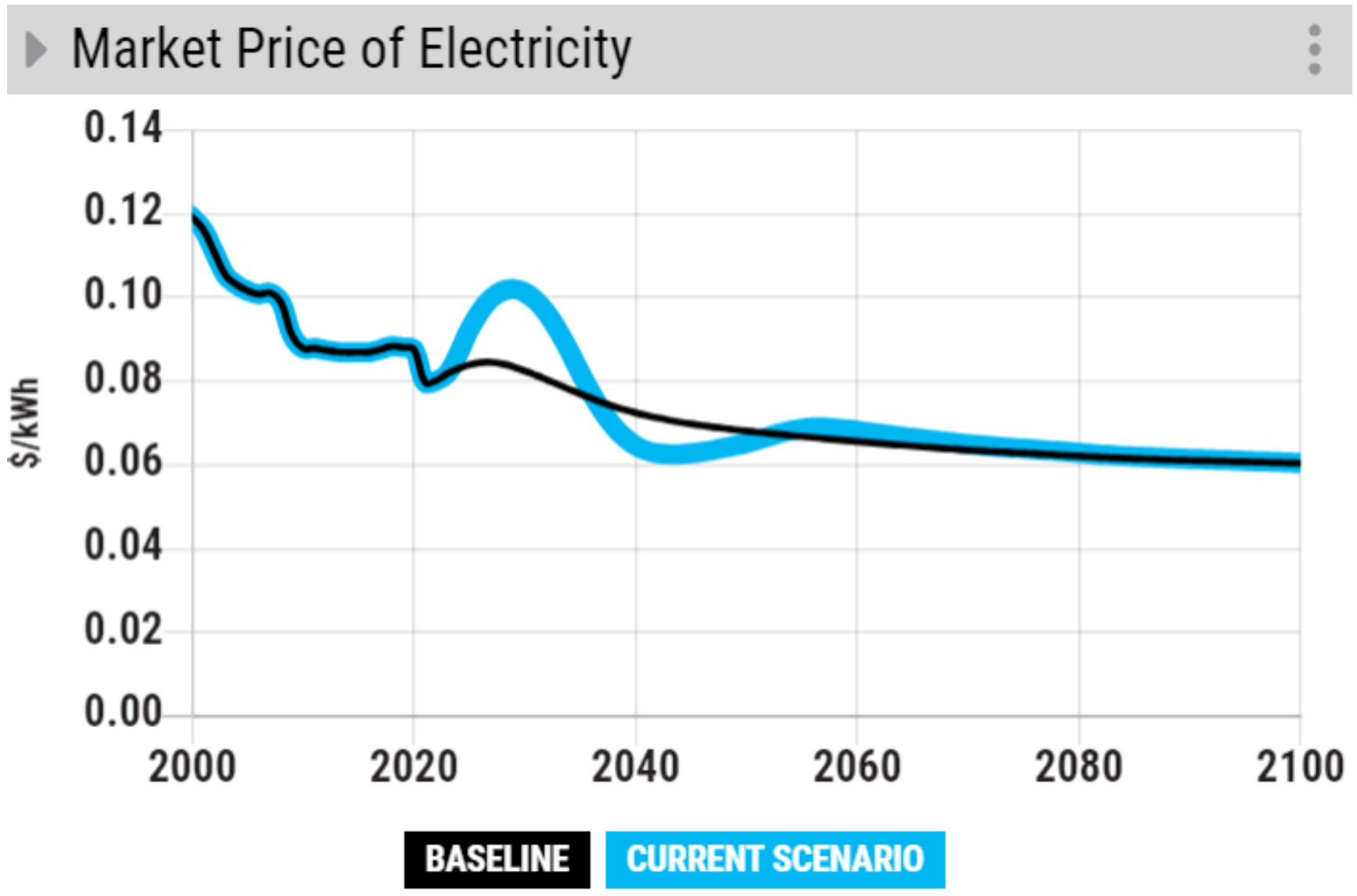

Figure 22: Market price of electricity with a gradual 30 year $250 carbon tax per ton emitted

Another important note would be the changing energy products procured by consumers. It can be inferred that the total price of electricity may remain the same, but the means of which consumers would obtain their electricity might be greener, and emit less carbon, given the stable pricing. By placing carbon dioxide emissions, and other less environmentally-friendly energy production methods at higher prices, by comparison, sustainable alternatives become more desirable, and thus mainstream, contributing to the notion of green finance (Lindenberg, 2014). As there is a higher demand for sustainable and renewable energy sources, there will be a greater number of suppliers willing to bring their products to the market, thus driving the price of sustainable energy alternatives down. As seen towards the tail end of both graphs, the price of electricity in total remains the same, however this production shift behind the scenes may have increasingly positive impacts in the long run, specifically on the composition of energy supplies across the world.

## 3.13 En-ROADS Solution: Optimized Government Policies

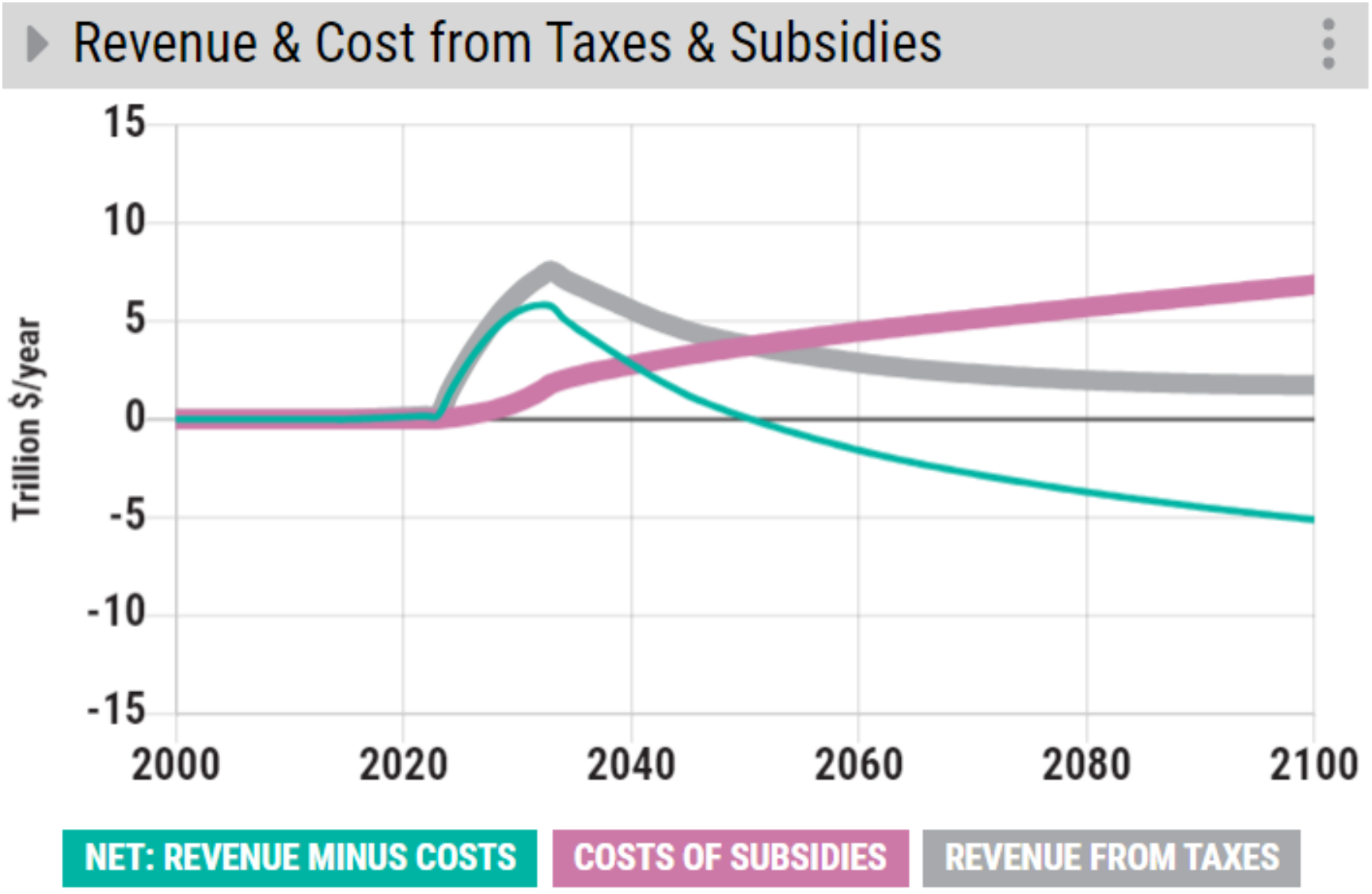

Figure 23: Revenue and cost from taxes and subsidies with heavy government involvement

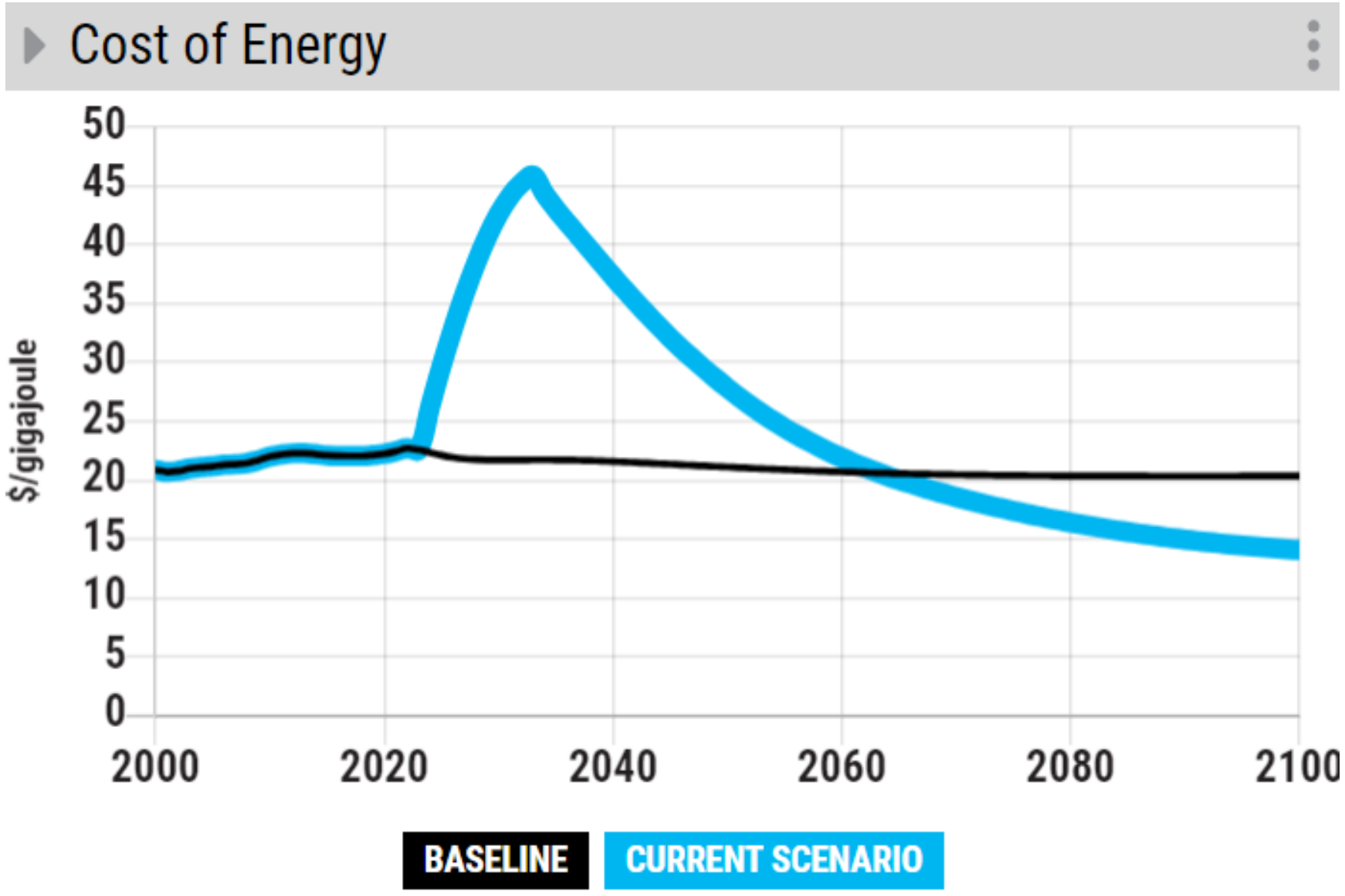

Figure 24: Cost of energy with heavy government involvement

Government budgets are an important factor to consider when policies are created, along with the long term impacts on the public and global price markets. To assess a situation with maximum government involvement, multiple factors and energy sources were taxed or subsidized. Coal, oil, natural gas, and carbon emissions were taxed highly, whilst renewables, biogenergy, and nuclear energy were subsidized heavily. This resulted with Figure 23, a revenue and cost graph derived from the taxes and subsidies imposed on these energy sources. As seen by the lighter pink line, the costs of subsidies exceeds the revenue from taxes in the year 2050. This also causes the government to go into a net loss for the remainder of the century. This illustrates the importance of a balanced budget, and strategic subsidies, as too many can drastically influence other parts of governance. On the other hand, this also produced a global temperature increase of just 2.3° Celsius, lower than any other scenario examined thus far. While this in itself is positive, the cost of energy is another upside, depending on the time scale. In the short run, under these policy-heavy conditions, the price of energy peaks in the short term, but dips under the baseline for the latter half of the century.

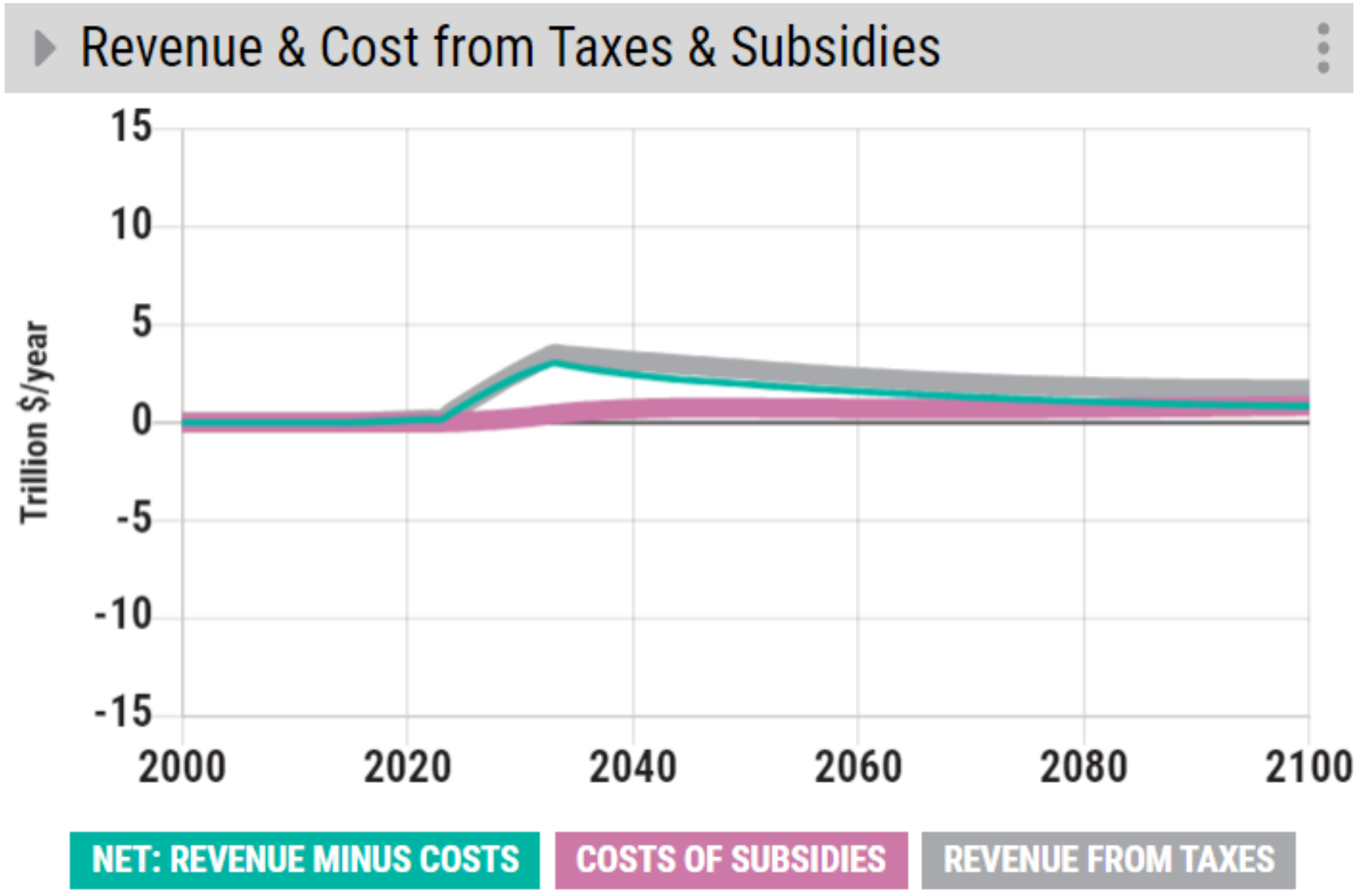

Figure 25: Revenue and cost from taxes and subsidies with optimized government involvement

In order to get a more realistic plan for subsidies and taxation sans long term governmental debt, a few of the features were optimized. The new scenario demonstrated in Figure 25, creates a situation where the temperature change is only a small increase from the previous example, to 2.5° Celsius. This scenario includes no taxation on carbon emissions, a $100 per TCE tax on coal, $85 per BOE tax on oil, $5 per MCF of natural gas, and subsidies of $8 per BOE of bioenergy, a moderate breakthrough on new zero carbon energy at double the price of coal, and $0.02 per kWh subsidy of nuclear and renewable energy. This represents a final, optimized scenario where significant climate progress is made, however minimal debt is incurred, and fewer markets are taxed to an extreme extent.

# 4. Conclusion and Policy Implications

Through analysis conducted on C-ROADS and En-ROADS, an optimized climate policy narrative was created, addressing the research gap of climate policies which incur economic growth.

First, the impact of sea level rise was observed, and it was found that there is an intense increase in flooding risk, especially in highly populated areas. Additionally, both ice caps (Antarctica and the Greenland Ice Sheet) had similar impacts on flood risk, meaning climate change policies must be enacted in most if not all locations. In terms of economic and population growth, economic growth was seen to have a larger effect on the total change in global temperature than population growth, thus emphasizing the importance of sustainable and responsible industrial decisions, which can be influenced through policy.

As climate sensitivity was found to have a significant impact on global temperature change, $CO_2$ emissions became more pertinent. Therefore, countries such as the US and China can have a significant impact on reducing emissions, whilst also decreasing $N_2O$ emissions, and mitigating health risk factors. China can also lead the way in afforestation efforts, due to previous success in this field.

In terms of energy supply, in the status quo, oil, coal and gas represent the largest shares of global energy supply. However, through moderate taxation enacted soon but with a gradual increase, the share of primary energy supply can be shifted almost entirely to renewables by 2100 whilst mitigating sudden spikes in energy costs. A moderate tax on carbon specifically can decrease the probability of an ice-free Arctic summer by half, which links back to lower flood risks.

Optimized government intervention can promote growth, whilst maintaining healthy public funds budgets. In the future, research may include investigations into the specific countries and industries for which subsidies and taxation would work best to retain the maximum climate benefit.

# 5. References

A. Betts, R. (2005). Integrated approaches to climate–crop modelling: needs and challenges. Philosophical Transactions of the Royal Society B: Biological Sciences, 360(1463), 2049-2065.

Anser, M. K., Godil, D. I., Khan, M. A., Nassani, A. A., Zaman, K., & Abro, M. M. Q. (2021). The impact of coal combustion, nitrous oxide emissions, and traffic emissions on COVID-19 cases: a Markov-switching approach. Environmental Science and Pollution Research, 28, 64882-64891.

Antwi-Agyei, P., Dougill, A. J., Agyekum, T. P., & Stringer, L. C. (2018). Alignment between nationally determined contributions and the sustainable development goals for West Africa. Climate Policy, 18(10), 1296-1312.

Bloom, D. E., Canning, D., & Sevilla, J. P. (2001). The effect of health on economic growth: theory and evidence. National Bureau of Economic Research Cambridge, Mass., USA.

Cho, K., Goldstein, B., Gounaridis, D., & Newell, J. P. (2021). Where does your guacamole come from? Detecting deforestation associated with the export of avocados from Mexico to the United States. Journal of Environmental Management, 278, 111482.

Creed, I. F., van Noordwijk, M., & Others. (2018). Forest and water on a changing planet: vulnerability, adaptation and governance opportunities. A global assessment report. IUFRO World Series, 38.

C-ROADS. (2023). Background — C-ROADS Technical Reference. Docs.climateinteractive.org. https://docs.climateinteractive.org/projects/c-roads-reference-guide/en/latest/pages/background.html

Debelle, G. (2019, March). Climate change and the economy. In Speech at public forum hosted by Centre for Policy Development, Sydney, Australia, March (Vol. 12).

De Paola, A., Ortolani, M., Lo Re, G., Anastasi, G., & Das, S. K. (2014). Intelligent management systems for energy efficiency in buildings: A survey. ACM Computing Surveys (CSUR), 47(1), 1–38.

En-ROADS. (2023). En-ROADS Technical Reference. Docs.climateinteractive.org. https://docs.climateinteractive.org/projects/en-roads-reference-guide/en/latest/index.html#index

Fankhauser, S., & Tol, R. S. J. (2005). On climate change and economic growth. Resource and Energy Economics, 27(1), 1–17.

Fiddaman. T. (1997) Feedback Complexity in Integrated Climate-Economy Models. Ph.D. Dissertation, Massachusetts Institute of Technology.

Filoso, S., Bezerra, M. O., Weiss, K. C. B., & Palmer, M. A. (2017). Impacts of forest restoration on water yield: A systematic review. PloS One, 12(8), e0183210.

Friedlingstein, P., Jones, M. W., O'sullivan, M., Andrew, R. M., Bakker, D. C., Hauck, J., ... & Zeng, J. (2022). Global carbon budget 2021. Earth System Science Data, 14(4), 1917-2005.

Gilland, B. (1995). World population, economic growth, and energy demand, 1990-2100: A review of projections. Population and Development Review, 507–539.

Gornitz, V., Couch, S., & Hartig, E. K. (2001). Impacts of sea level rise in the New York City metropolitan area. Global and Planetary Change, 32(1), 61-88.

Hensel, M., Bryan, J., McCarthy, C., McNeal, K. S., Norfles, N., Rath, K., & Rooney-Varga, J. N. (2023). Participatory approaches enhance a sense of urgency and collective efficacy about climate change: Qualitative evidence from the world climate simulation. Journal of Geoscience Education, 71(2), 177-191.

Holz, C., Siegel, L. S., Johnston, E., Jones, A. P., & Sterman, J. (2018). Ratcheting ambition to limit warming to 1.5 C–trade-offs between emission reductions and carbon dioxide removal. Environmental research letters, 13(6), 064028.

Jevrejeva, S., Grinsted, A., & Moore, J. C. (2014). Upper limit for sea level projections by 2100. Environmental Research Letters, 9(10), 104008.

Kaplan, J. O., Krumhardt, K. M., & Zimmermann, N. (2009). The prehistoric and preindustrial deforestation of Europe. Quaternary Science Reviews, 28(27–28), 3016–3034.

Kapmeier, F., Greenspan, A., Jones, A., & Sterman, J. (2021). Science-based analysis for climate action: how HSBC Bank uses the En-ROADS climate policy simulation. System dynamics review: the journal of the System Dynamics Society, 37(4), 333-352.

Kumar, R. (2022). US-China Trade War: Impact on Sustainable Development in Developing Nations with Particular Reference to South Asia.

Lake, I. R., Hooper, L., Abdelhamid, A., Bentham, G., Boxall, A. B., Draper, A., ... & Waldron, K. W. (2012). Climate change and food security: health impacts in developed countries. Environmental health perspectives, 120(11), 1520-1526.

Liguo, X., Ahmad, M., & Khattak, S. I. (2022). Impact of innovation in marine energy generation, distribution, or transmission-related technologies on carbon dioxide emissions in the United States. Renewable and Sustainable Energy Reviews, 159, 112225.

Lindenberg, N. (2014). Definition of green finance. In Definition of Green Finance: Lindenberg, Nannette. [Sl]: SSRN.

Liu, N., Salauddin, M., Yeganeh-Bakhtiari, A., Pearson, J., & Abolfathi, S. (2022, September). The Impact of Eco-retrofitting on Coastal Resilience Enhancement–A Physical Modelling Study. In IOP Conference Series: Earth and Environmental Science (Vol. 1072, No. 1, p. 012005). IOP Publishing.

McNicholas, R. (2023). Climate Change Proposal: Coupling Equity and Scientific Rigor in Facing Global Warming.

Meo, M. S., & Abd Karim, M. Z. (2022). The role of green finance in reducing CO2 emissions: An empirical analysis. Borsa Istanbul Review, 22(1), 169-178.

Oeschger, H., Siegenthaler, U., et al. (1975) A Box Diffusion Model to Study the Carbon Dioxide Exchange in Nature. Tellus, 27(2):167-192.

Peres, C. A., Campos-Silva, J., & Ritter, C. D. (2023). Environmental policy at a critical junction in the Brazilian Amazon. Trends in Ecology & Evolution.

Pfaff, A. S. P. (2000). From deforestation to reforestation in New England, United States. World Forests from Deforestation to Transition?, 67–82.

Reddy, C. S., Dutta, K., & Jha, C. S. (2013). Analysing the gross and net deforestation rates in India. Current Science, 1492–1500.

Report: State and Trends of Carbon Pricing --- worldbank.org. (n.d.). Retrieved from https://www.worldbank.org/en/news/press-release/2022/05/24/global-carbon-pricing-generates-record-84-billion-in-revenue

Rooney-Varga, J. N., Kapmeier, F., Sterman, J. D., Jones, A. P., Putko, M., & Rath, K. (2020). The climate action simulation. Simulation & Gaming, 51(2), 114–140.

Sanchez, T. W., Shumway, H., Gordner, T., & Lim, T. (2023). The prospects of artificial intelligence in urban planning. International Journal of Urban Sciences, 27(2), 179-194.

Siegel, L. S., Homer, J., Fiddaman, T., McCauley, S., Franck, T., Sawin, E., … Interactive, C. (2018). En-roads simulator reference guide. Technical Report.

Sigmond, M., Fyfe, J. C., & Swart, N. C. (2018). Ice-free Arctic projections under the Paris Agreement. Nature Climate Change, 8(5), 404–408.

Sterman, J., Fiddaman, T., Franck, T. R., Jones, A., McCauley, S., Rice, P., … Siegel, L. (2012). Climate interactive: the C-ROADS climate policy model.

Stocker, T. (2014). Climate change 2013: the physical science basis: Working Group I contribution to the Fifth assessment report of the Intergovernmental Panel on Climate Change. Cambridge university press.

Stunkel, L. & Tucker, J. (2020). China's Afforestation Efforts: Not Seeing the Forest for the Trees? --- isdp.se. Retrieved from https://isdp.se/china-afforestation-efforts/

United Nations. (2022). Climate Change - United Nations Sustainable Development. Sustainable Development Goals; United Nations. https://www.un.org/sustainabledevelopment/climate-change/

Verhoef, P. C., Noordhoff, C. S., & Sloot, L. (2023). Reflections and predictions on effects of COVID-19 pandemic on retailing. Journal of Service Management, 34(2), 274-293.

Vermeer, M. and S. Rahmstorf. 2009. Global sea level linked to global temperature. Proc of the Nat Acad of Sci. 106(51):21527-21532. www.pnas.org/cgi/doi/10.1073/pnas.0907765106.

Vitale, C., Meijerink, S., & Moccia, F. D. (2023). Urban flood resilience, a multi-level institutional analysis of planning practices in the Metropolitan City of Naples. Journal of Environmental Planning and Management, 66(4), 813-835.

Winton, M. (2006). Amplified Arctic climate change: What does surface albedo feedback have to do with it? Geophysical Research Letters, 33(3).

Wyatt, S. N., Sullivan‐Watts, B. K., Watts, D. R., & Sacks, L. A. (2022). Facilitating Climate Change Action in the Ocean Sciences Using the Interactive Computer Model En‐ROADS.

Zahar. (2019, May 22). Climate Simulation: Role Playing Exercise that Feels Like the Real Thing. Www.climateinteractive.org. https://www.climateinteractive.org/blog/climate-simulation-role-playing-exercise-that-feels-like-the-real-thing/

Zhang, Yuxing, & Song, C. (2006). Impacts of afforestation, deforestation, and reforestation on forest cover in China from 1949 to 2003. Journal of Forestry, 104(7), 383–387.

Zhang, M., Liu, N., Harper, R., Li, Q., Liu, K., Wei, X., … Liu, S. (2017). A global review on hydrological responses to forest change across multiple spatial scales: Importance of scale, climate, forest type and hydrological regime. Journal of Hydrology, 546, 44–59.

Zhang, M., & Wei, X. (2021). Deforestation, forestation, and water supply. Science, 371(6533), 990–991.

Zhang, Yajuan, Zhang, L., Wang, H., Wang, Y., Ding, J., Shen, J., … Li, S. (2022). Reconstructing deforestation patterns in China from 2000 to 2019. Ecological Modelling, 465, 109874.

# 6. Appendix

A. Notable “developed countries” (for calculation purposes only): Canada, Japan, United Kingdom, Australia
   Notable “developing countries” (for calculation purposes only): South Africa, Mexico, Brazil, and Indonesia

B. India’s $N_2O$ Emissions (Megatons $N_2O$/year) - Following India decreasing total emissions by 10% starting in 2050

| **Year** | **Baseline (Mts)** | **Emissions Reduction Scenario (Mts)** |
|---:|---:|---:|
| 2000 | 5.77 | 5.77 |
| 2001 | 5.77 | 5.77 |
| 2002 | 5.9 | 5.9 |
| 2003 | 5.93 | 5.93 |
| 2004 | 6.16 | 6.16 |
| 2005 | 6.18 | 6.18 |
| 2006 | 6.27 | 6.27 |
| 2007 | 6.42 | 6.42 |
| 2008 | 6.32 | 6.32 |
| 2009 | 6.26 | 6.26 |
| 2010 | 6.35 | 6.35 |
| 2011 | 6.6 | 6.6 |
| 2012 | 6.66 | 6.66 |
| 2013 | 6.6 | 6.6 |
| 2014 | 6.69 | 6.69 |
| 2015 | 6.71 | 6.71 |
| 2016 | 6.8 | 6.8 |
| 2017 | 6.92 | 6.92 |
| 2018 | 6.93 | 6.93 |
| 2019 | 7 | 7 |
| 2020 | 6.96 | 6.96 |
| 2021 | 7.09 | 7.09 |
| 2022 | 7.18 | 7.18 |
| 2023 | 7.27 | 7.27 |
| 2024 | 7.36 | 7.36 |
| 2025 | 7.45 | 7.45 |
| 2026 | 7.55 | 7.55 |
| 2027 | 7.64 | 7.64 |
| 2028 | 7.73 | 7.73 |
| 2029 | 7.83 | 7.83 |
| 2030 | 7.92 | 7.92 |
| 2031 | 8.01 | 8.01 |
| 2032 | 8.1 | 8.1 |
| 2033 | 8.2 | 8.2 |
| 2034 | 8.29 | 8.29 |
| 2035 | 8.38 | 8.38 |
| 2036 | 8.46 | 8.46 |
| 2037 | 8.55 | 8.55 |
| 2038 | 8.64 | 8.64 |
| 2039 | 8.72 | 8.72 |
| 2040 | 8.81 | 8.81 |
| 2041 | 8.89 | 8.89 |
| 2042 | 8.97 | 8.97 |
| 2043 | 9.05 | 9.05 |
| 2044 | 9.13 | 9.13 |
| 2045 | 9.2 | 9.2 |
| 2046 | 9.27 | 9.27 |
| 2047 | 9.35 | 9.35 |
| 2048 | 9.42 | 9.42 |
| 2049 | 9.49 | 9.49 |
| 2050 | 9.55 | 9.55 |
| 2051 | 9.62 | 9.37 |
| 2052 | 9.68 | 9.21 |
| 2053 | 9.74 | 9.07 |
| 2054 | 9.81 | 8.95 |
| 2055 | 9.86 | 8.84 |
| 2056 | 9.92 | 8.74 |
| 2057 | 9.98 | 8.66 |
| 2058 | 10.03 | 8.59 |
| 2059 | 10.09 | 8.53 |
| 2060 | 10.14 | 8.47 |
| 2061 | 10.19 | 8.43 |
| 2062 | 10.24 | 8.39 |
| 2063 | 10.29 | 8.36 |
| 2064 | 10.33 | 8.33 |
| 2065 | 10.38 | 8.31 |
| 2066 | 10.42 | 8.3 |
| 2067 | 10.47 | 8.29 |
| 2068 | 10.51 | 8.28 |
| 2069 | 10.55 | 8.27 |
| 2070 | 10.59 | 8.27 |
| 2071 | 10.63 | 8.27 |
| 2072 | 10.67 | 8.28 |
| 2073 | 10.71 | 8.28 |
| 2074 | 10.74 | 8.29 |
| 2075 | 10.77 | 8.3 |
| 2076 | 10.81 | 8.3 |
| 2077 | 10.84 | 8.31 |
| 2078 | 10.87 | 8.32 |
| 2079 | 10.9 | 8.34 |
| 2080 | 10.92 | 8.35 |
| 2081 | 10.95 | 8.36 |
| 2082 | 10.98 | 8.37 |
| 2083 | 11 | 8.38 |
| 2084 | 11.02 | 8.4 |
| 2085 | 11.05 | 8.41 |
| 2086 | 11.07 | 8.42 |
| 2087 | 11.09 | 8.44 |
| 2088 | 11.11 | 8.45 |
| 2089 | 11.13 | 8.46 |
| 2090 | 11.15 | 8.48 |
| 2091 | 11.17 | 8.49 |
| 2092 | 11.19 | 8.51 |
| 2093 | 11.21 | 8.52 |
| 2094 | 11.23 | 8.54 |
| 2095 | 11.25 | 8.56 |
| 2096 | 11.27 | 8.57 |
| 2097 | 11.29 | 8.59 |
| 2098 | 11.31 | 8.61 |
| 2099 | 11.33 | 8.63 |
| 2100 | 11.35 | 8.65 |